\documentclass[12pt, draftclsnofoot, onecolumn]{IEEEtran}
\usepackage{amsmath,amsfonts}
\usepackage{algorithmic}
\usepackage{array}
\ifCLASSOPTIONcompsoc
    \usepackage[caption=false, font=normalsize, labelfont=sf, textfont=sf]{subfig}
\else
\usepackage[caption=false, font=footnotesize]{subfig}
\fi
\usepackage{textcomp}
\usepackage{stfloats}
\usepackage{url}
\usepackage{cite}
\usepackage{verbatim}
\usepackage{graphicx}
\usepackage[usenames]{color}
\usepackage{xcolor}
\usepackage{color}
\def\BibTeX{{\rm B\kern-.05em{\sc i\kern-.025em b}\kern-.08em
    T\kern-.1667em\lower.7ex\hbox{E}\kern-.125emX}}
\usepackage{balance}
\begin{document}
\title{CSI-fingerprinting Indoor Localization via Attention-Augmented Residual Convolutional Neural Network}
\author{Bowen Zhang$^{\ast}$~\IEEEmembership{Student Member,~IEEE}, Houssem Sifaou~\IEEEmembership{Member,~IEEE}, Geoffrey Ye Li~\IEEEmembership{Fellow,~IEEE}
\thanks{This work was sponsored by the donation from Huawei Technologies.} \thanks{$^{\ast}$Corresponding author}
\thanks{
Bowen Zhang, Houssem Sifaou and Geoffrey Ye Li are with the Department of Electrical and Electronic Engineering, Imperial College London, London SW7 2AZ, U.K. (e-mail: k.zhang21@imperial.ac.uk; h.sifaou@imperial.ac.uk; geoffrey.li@imperial.ac.uk)}}

%\markboth{Journal of \LaTeX\ Class Files,~Vol.~18, No.~9, September~2020}%
%{How to Use the IEEEtran \LaTeX \ Templates}

\maketitle

\begin{abstract}
Deep learning has been widely adopted for channel state information (CSI)-fingerprinting indoor localization systems. These systems usually consist of two main parts, \textit{i}.\textit{e}., a positioning network that learns the mapping from high-dimensional CSI to physical locations and a tracking system that utilizes historical CSI to reduce the positioning error. This paper presents a new localization system with high accuracy and generality. On the one hand, the receptive field of the existing convolutional neural network (CNN)-fingerprinting positioning networks is limited, restricting their performance as useful information in CSI is not explored thoroughly. As a solution, we propose a novel attention-augmented residual CNN to utilize the local information and global context in CSI exhaustively. On the other hand, considering the generality of a tracking system, we decouple the tracking system from the CSI environments so that one tracking system for all environments becomes possible. Specifically, we remodel the tracking problem as a denoising task and solve it with deep trajectory prior. Furthermore, we investigate how the precision difference of inertial measurement units will adversely affect the tracking performance and adopt plug-and-play to solve the precision difference problem. Experiments show the superiority of our methods over existing approaches in performance and generality improvement.   
\end{abstract}


\section{Introduction}
\IEEEPARstart{I}{n} recent decades, accurate positioning is receiving increasing interests as a key enabler for many location-based services, such as navigation, smart robots, and the internet of things. However, the widely-used
global positioning system is not applicable in indoor scenarios since the line-of-sight (LOS) propagation between the mobile terminals (MTs) and the
satellites can be blocked. Therefore, many indoor positioning techniques have been proposed as alternative solutions.

Among these alternatives, channel state information (CSI)-based fingerprinting localization has achieved increasing attention for its simplicity, broad applicability, and reliability~\cite{xiao2012fifs,wang2016csi,
wang2018wifi,wang2018deep,zhang2020indoor}. These CSI-fingerprinting localization methods usually consist of two phases: 1) an offline phase, where the CSI is measured at some spatially-distributed reference points (RPs) and a localization algorithm is developed based on the collected CSI-location dataset, and 2) an online testing phase, in which the developed localization algorithm is applied to predict the location of MT using real-time CSI. CSI-fingerprinting localization does not need to model the wireless channel explicitly and works for non-line-of-sight (NLOS) scenarios.

%Since the channel state is determined by the wireless propagation environment around the MT, the CSI-based positioning approach transforms the localization problem into a pattern recognition problem. Specifically, it consists of a data collection stage where CSI measurements are taken at spatially-distributed reference points (RPs), an algorithm development stage where the CSI at different RPs are used to optimize the matching algorithm, and an online estimation stage where the MT location is estimated using the matching algorithm. 

Inspired by the great success of deep learning, some works treat CSI as images and use convolutional neural networks (CNNs) to learn the mapping from CSI measurements to two-dimensional terminal coordinates~\cite{vieira2017deep,sun2019fingerprint,cerar2021improving,
chin2020intelligent}. After training on the CSI-location pairs collected at the RPs, these CNN-based methods achieve higher positioning accuracy than the traditional probabilistic methods~\cite{xiao2012fifs}. 

Although the existing methods have successfully introduced deep CNN into the CSI-based fingerprinting indoor localization problem, we find that using a CNN architecture with a larger receptive field (RF) can improve the performance. Unlike in fully connected networks (FCNs), where the value of each pixel depends on the
entire input, the value of a pixel in a CNN layer is only affected by the inputs in a region, which is defined as the RF of that CNN layer~\cite{luo2016understanding}. As each pixel in one layer cannot see the input information outside its RF, it is essential to ensure the RF is large enough to cover all the relevant information for accurate decision making. For CSI-fingerprinting indoor positioning, the requirement for a large RF is significant as we need to leverage the channel responses on multiple non-neighbouring frequency bands and antennas to alleviate the influence of environmental noise~\cite{tong2020csi} and fading~\cite{xiao2012fifs}.
 
Besides the positioning system, a tracking system is usually
built to improve the localization accuracy for moving targets. As the motion patterns in an indoor environment are relatively fixed due to environmental restrictions, the tracking system can use the historical information on the trajectory to enhance the location estimation for the current timestamp~\cite{bai2019dl}. In the context of indoor tracking, a generic approach is to take the results from a positioning system as measurements and track the state of the target, including position, velocity, and acceleration, over time via Bayesian estimators~\cite{shi2018accurate,choi2022sensor
,zhou2019freetrack}. Nevertheless, the success of these parametric model-based methods relies on the accuracy of system models~\cite{ge2016performance}, which
are difficult to estimate in many cases. %Existing CSI-based indoor tracking studies based on Bayesian estimators usually assume the target has a nearly-constant velocity for simplicity~\cite{shi2018accurate,choi2022sensor,zhou2019freetrack}.

To address the above problem, integrating the sensory data from an inertial measurement unit (IMU) into a tracking system is a popular direction~\cite{belmonte2019swiblux,li2016fine}. By using built-in IMUs, such as acceleration sensors and gyroscopes, step length and orientation can be estimated through the pedestrian dead reckoning (PDR) approach~\cite{ho2016step}, with which a more realistic system model can be built.

Some learning-based tracking~\cite{hoang2020cnn,zhang2021deep,bai2019dl} or IMU-aided tracking methods~\cite{xie2020learning,chen2021data} have recently been proposed for indoor environments. By exploring the correlation of CSI measurements or IMU data along a trajectory, these methods achieve better performance than stationary positioning systems
and Bayesian filtering tracking systems. 

Despite the performance improvement, the generality of these learning-based tracking systems can be improved. Expressly, most existing tracking methods assume that the tracking network is trained for each indoor environment. At the same time, this is not easy as training a tracking system requires collecting precise positions and CSI/IMU measurements for a continuously moving target at discrete time instances, which are difficult to measure precisely without enough time and workforce~\cite{xie2020learning}. Also, existing works mainly focus on one type of IMU and do not consider the precision difference between different types of IMU. We find it impossible for a single network to handle IMU measurements with varying precision while training a specific model for each type is a huge burden. Therefore, it is necessary for a universal tracking system to be used across indoor environments and compatible for IMU equipment with arbitrary precision. 

%Despite the performance improvement, the generality of these learning-based tracking systems can be improved. A tracking system is simultaneously influenced by a series of factors, such as the wireless propagation environment, the accuracy of the positioning system, the precision of IMU equipment, and the motion patterns of the MT, while existing learning-based methods are mainly trained and tested under a specific condition and do not take into account the changes of these factors. 

%Due to the complicated data collection stage, a universal tracking system is vital to put the tracking system into practice. Unlike collecting CSI at fixed RPs in a positioning system, learning a tracking system requires recording the coordinates and the CSI/IMU measurements for the continuous trajectories a moving target has made, which are difficult to measure precisely
%without enough time and workforce~\cite{xie2020learning}. Therefore, it is crucial to ensure the tracking system can be used across indoor environments for any IMU equipment without recollecting a joint CSI/IMU-trajectory dataset. 

In this paper, we consider both CSI-fingerprinting positioning and
tracking. For positioning, we will enhance the feature extraction process of the positioning network by increasing its RF. Specifically, we first enlarge the RF by increasing the depth of our network with stacked residual blocks~\cite{lim2017enhanced}. Then, we extract global context from CSI through an attention-augmented convolutional operation~\cite{bello2019attention}. Besides, we re-model tracking as a denoising task, where the trajectory generated by a positioning system is a noisy observation of the true trajectory, and explore the inherent properties of indoor trajectories to mitigate the noise. These properties are expressed by a trajectory prior function in our work. 
%due to the walking habits and indoor environment restrictions, the shape of the trajectory for indoor environments should have some universal properties. For instance, most of the time, people will walk along a straight path to get to the destination as soon as possible. Also, some indoor environments, like corridors and hallways, are constructed straight~\cite{ilyas2016drift}. In this case, the trajectories should be a line. Therefore, 
%we can constrain the solution space for the denoising task and reduce the localization error by using this sort of regularization, which is called trajectory prior in our work. 
As the trajectory prior function is independent of the wireless propagation environments and positioning systems, our tracking system has better generality. %As for the definition of trajectory prior, this paper learns it from data instead of defining it manually. 
    
Moreover, for IMU-aided tracking systems, we consider the precision difference between different IMUs~\cite{yan2019performance}. We adopt the idea of plug-and-play (PnP)~\cite{zhang2017learning} 
to integrate the IMU data into a learning-based tracking system in a model-based way, so that the precision difference problem can be handled by tuning the trade-off parameters in the model.

Our main contributions can be summarized as follows:
\begin{enumerate}
\item{To improve the positioning performance, we develop a novel attention-augmented residual convolutional neural network (AAResCNN) for CSI-based fingerprinting indoor positioning. Experiments on publicly-released datasets show that our network can achieve higher localization accuracy than state-of-the-art (SOTA) CNN-based methods.} 
\item{We propose a universal tracking method by training a set of Gaussian Denoiser\textcolor{blue}{s} based on the deep trajectory prior function. Simulation results verify the generality of our tracking system for different environments.
%which relieves us from the need of collecting a CSI-trajectory joint dataset for each wireless environment and retraining the tracking network when environment changes. 
}
\item{We use PnP to integrate IMU sensory data into the tracking system. Experiments on simulated datasets show that our methods perform better than learning-based and Bayesian filtering-based methods when the signal-to-noise-ratio (SNR) of IMU sensory data changes.}
\end{enumerate}
In the remainder of the paper, related works are reviewed in
Section~\ref{review}. In Section~\ref{overall}, we present the overall architecture of our system, followed by our motivations and contributions. Sections~\ref{position},~\ref{tracking} and~\ref{imu tracking} introduce our  positioning, tracking, and IMU-aided tracking methods, respectively. The experimental setups and evaluation results are presented in Section~\ref{experiments}. Finally, conclusions of our work are shown in Section~\ref{conclusion}.

\section{Related Works}
\label{review}
Many studies have been conducted for CSI-based fingerprinting indoor localization. This section divides the existing approaches into three categories, i.e., positioning, tracking and IMU-aided tracking, and presents part of representative works for each category. Furthermore, this section describes the difference between our work and the existing ones.
\subsection{CSI-fingerprinting Positioning}
CSI-fingerprinting indoor positioning systems have achieved wide attentions in recent years. In 2012, Xiao et al.~\cite{xiao2012fifs} proposed FIFS using the summation of power across frequency bands in CSI as fingerprints and achieved higher accuracy than traditional received signal strength indicator (RSSI)-based approach. In~\cite{jaffe2014single}, channel responses on multiple antennas were aggregated for CSI-based fingerprinting indoor positioning. Afterwards, exploring channel information in both frequency and spatial domain becomes the mainstream.

With the development of machine learning, many learning-based approaches have been proposed to improve localization accuracy using CSI fingerprints~\cite{wang2016csi,wang2018wifi,wang2018deep
,zhang2020indoor},\cite{zhu2021bls}. For example, in~\cite{wang2016csi}, restricted Bolzmann machines (RBMs) were used to store the CSI-fingerprints as trainable weights for following matching algorithms. Long short-term
memory (LSTM) was adopted in~\cite{zhang2020indoor} to explore the correlation of CSI features over time. X. Wang et al. in~\cite{wang2018deep} constructed a set of CSI images from received packets and used deep CNN
for feature extraction and location estimation. In~\cite{zhu2021bls}, a Broad Learning System is introduced to reduce the training time of positioning systems.

The methods mentioned above all collect multiple received packets for one position to suppress channel fluctuation and noise. However, it is hard to collect multiple packets in a short time when tracking a moving target. Therefore, positioning with a single CSI sample is required. Several CNN-based methods that take a single CSI sample as input have been proposed~\cite{vieira2017deep,sun2019fingerprint,cerar2021improving,
chin2020intelligent,wu2021learning}. In this paper, we propose a novel CNN
architecture that leads to better positioning accuracy.
%\textit{et} \textit{al}. 
\subsection{Tracking with CSI-fingerprinting}
%Instead of regarding the continuous CSI measurements along a trajectory as mutually independent observations, CSI-based tracking aims to enhance the positioning accuracy by using historical trajectory information. 
Bayesian filtering has been widely adopted for tracking tasks with CSI-fingerprinting. For example, S. Shi et al. in~\cite{shi2018accurate} exploited the Kalman filter (KF) to continuously track the trajectory of a moving target using the location results from a CSI-fingerprinting positioning system. In~\cite{choi2022sensor}, enhanced Kalman filter (EKF) was applied to recover the trajectory from CSI measurements. Particle filter (PF)-based tracking methods were also proposed in~\cite{shi2018accurate,zhou2019freetrack}.

LSTM-based tracking methods have also been proposed for indoor environments recently. For instance, in~\cite{hoang2020cnn,zhang2021deep}, continuous CSI measurements of trajectories were first represented as deep features via a CNN network, and then these features were fed into an LSTM network for tracking purposes. However, these learning-based tracking methods are designed under a specific environment and retraining is required when indoor environment changes. In this paper, we will design a universal tracking network from a trajectory dataset alone without using CSI measurements.

\subsection{IMU-aided Tracking}
As shown in~\cite{belmonte2019swiblux,li2016fine,
chen2021data}, fusing the IMU sensor data with wireless signals can help reduce the positioning error. %A generic approach to use IMU data is to use the estimated step length and orientation from the PDR approach to construct a more reliable system model for Bayesian estimators.
Recently, there have been some works to explore the functionality of IMU data in a learning-based tracking system. For example, A. Xie et al. in~\cite{xie2020learning} used the IMU sensor data to learn the transition function in a state-space model. In~\cite{chen2021data}, the trajectory recovered by the CSI measurements was refined by the trajectory estimated through the PDR approach via back-propagation. Nevertheless, the influence of IMU precision on the tracking system has not been discussed.
\section{System Architecture and Contributions}
\label{overall}
In this section, we will first present the overall architecture of the proposed localization system, and then introduce the motivations and contributions for each module in our system.   
\subsection{System Architecture}
\begin{figure}[t]
\centering
\includegraphics[scale=0.75]{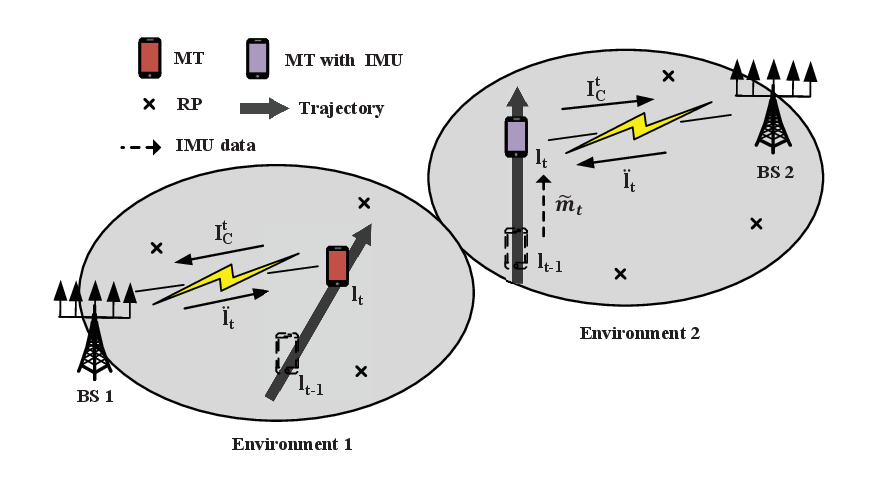}
\caption{The system architecture overview.}
\label{fig:system}
\end{figure}
The overall architecture of our localization system is shown in Fig~\ref{fig:system}. 
%from both the base station (BS) side and the MT side. Specifically, as shown in Fig~\ref{fig:system}, we assume the positioning system is deployed at the BS side and the tracking system at the MT side. 
Suppose a single base station (BS) is providing localization services for the terminals in an indoor environment. When the BS receives a localization request from a terminal in its coverage at time $t$, the BS first measures the CSI from the received packets. Assuming the BS works on a multiple-input multiple-output (MIMO) orthogonal frequency-division multiplexing (OFDM) system with $A$ receive antennas and $S$ sub-carriers, the measured CSI can be represented by a channel response matrix $I_{C}^{t} \in \mathcal{R}^{A\times S\times 2}$, where $2$ denotes the real part and imaginary part of the complex channel response, $A$ and $S$ denote the spatial dimension and the frequency dimension of CSI, respectively. $A\times S$ denotes the spatial size of CSI. 
\iffalse
Furthermore, if the antennas are placed as a uniform linear array (ULA), the channel response on the $a$-th antenna and $s$-th sub-carrier can be expressed as~\cite{sun2019fingerprint},
\begin{equation}
[I_{C}^{t}]_{a,s}=\sum_{p=1}^{P}c_{p}e^{-j2\pi\frac{ad\cos(\phi_{p})}{\lambda_{c}}}e^{-j2\pi f_{s}\tau_{p}},
\end{equation}
where $P$ denotes the number of distinguishable paths between the MT and the BS; $c_{p}$, $\phi_{p}$, and $\tau_{p}$ denote the complex channel coefficient, the angle of arrival, and the time delay for the $p$-th path, respectively; $d$ denotes the spatial distance between the adjacent antennas, $\lambda_{c}$ denotes the wavelength of the center frequency, and $f_{s}$ denotes the frequency of the $s$-th sub-carrier. 
\fi

\begin{figure}[t]
\centering
\subfloat[\centering BS side]{
\includegraphics[width=0.23\textwidth]{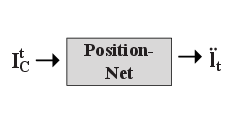}
}
\hfill
\subfloat[\centering MT side]{
\includegraphics[width=0.34\textwidth]{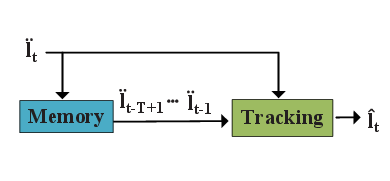}
}
\hfill
\subfloat[\centering MT side with IMU]{
\includegraphics[width=0.35\textwidth]{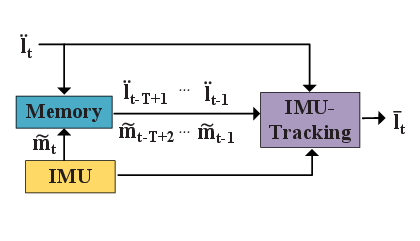}
}
\\
\caption{The deployment details at BS side and MS side.}
\label{fig:BS-MT}
\end{figure}
The BS then estimates the location of the MT using a CSI-fingerprinting positioning network, as shown in Fig~\ref{fig:BS-MT} (a),
\begin{equation}
\ddot{l}_{t}=\text{Position-Net}(I_{C}^{t}),
\end{equation}
where $\text{Position-Net}(\cdot)$ denotes the CSI-fingerprinting positioning network. %Note that before putting a positioning network into use, the parameters of the network should be trained on a CSI-location dataset. To construct the dataset, CSI is measured at some RPs in the indoor environment with known locations. In general, both the number of RPs and the positioning network architecture will affect the positioning system's accuracy. Also, as the CSI is environment-specific, the training process is repeated for each indoor environment, and the positioning network deployed on each BS has different parameters. 
Afterwards, the BS transmits the estimated localization result to the MT through the communication link.

When the MT receives the location estimation, $\ddot{l}_{t}$, from the BS, it will exploit historical trajectory information to reduce the positioning error via a tracking system iteratively~\cite{shi2018accurate}. Specifically, it  will first search its memory and generate a trajectory that consists of $T$ time steps,
\begin{equation}
\ddot{L}_{t}=\{\ddot{l}_{t-T+1},\ddot{l}_{t-T+2},\cdots,\ddot{l}_{t-1},\ddot{l}_{t}\},
\end{equation}
where $\ddot{L}_{t}$ denotes the trajectory at time $t$. The trajectory is then refined by a tracking system, as shown in Fig~\ref{fig:BS-MT} (b),
\begin{equation}
\hat{L}_{t}=\text{Tracking}(\ddot{L}_{t}),
\label{equ:track}
\end{equation}
where $\text{Tracking}(\cdot)$ denotes the tracking system. $\hat{L}_{t}=\{\hat{l}_{t-T+1},\hat{l}_{t-T+2},\cdots,\hat{l}_{t-1},\hat{l}_{t}\}$ denotes the refined trajectory after considering historical trajectory information. %To build the tracking system, an indoor trajectory dataset is needed in our work. The dataset should include most of the trajectory types in indoor environments, which can be simulated theoretically or collected by a low-cost autonomous mobile terminal. 
At last, the MT uses $\hat{l}_{t}$ as its location at time $t$. %Note that as the MT will move from one indoor environment to another, therefore the tracking system deployed at MT side should have a good generality. 
%As the MT will move from one indoor environment to another, the tracking system should have a good generalization capacity.

Most of the MTs nowadays are equipped with IMUs, such as accelerometers, gyroscopes, and magnetometers~\cite{ho2016step}. The MTs can utilize the IMU measurements $\tilde{M}_{t}=\{\tilde{m}_{t-T+2},\cdots,\tilde{m}_{t-1},\tilde{m}_{t}\}$ for tracking, where $\tilde{m}_{t}=[\tilde{r}_{t},\tilde{\theta}_{t}]$ denotes the estimated moving distance and direction from time $t-1$ to time $t$, respectively~\cite{belmonte2019swiblux}. The number of IMU meaurements is equal to the number of time intervals. Therefore, $\tilde{M}_{t}$ has a length of $T-1$ when trajectory $\ddot{L}_{t}$ lasts $T$ time steps. The trajectory refined by the IMU-aided tracking system can be represented as
\begin{equation}
\bar{L}_{t}=\text{IMU-Tracking}(\ddot{L}_{t},\tilde{M}_{t}),
\end{equation}
where $\text{IMU-Tracking}(\cdot)$ denotes the IMU-assisted tracking system. $\bar{L}_{t}=\{\bar{l}_{t-T+1},\bar{l}_{t-T+2},\cdots,\bar{l}_{t-1},$ $\bar{l}_{t}\}$ denotes the refined trajectory considering both the positioning network and IMUs, as shown in Fig~\ref{fig:BS-MT} (c). Similarly, the MT uses $\bar{l}_{t}$ as the final result.

In our system, the tracking system is deployed at the MT side, rather than the BS side. This means the historical trajectory information does not need to be exchanged between BSs when MT moves across environments, increasing the efficiency of the system.
\subsection{CSI-fingerprinting positioning network}
CSI-fingerprinting positioning network needs to be trained on a CSI-location dataset to work properly. This data collection task is time-consuming and laborious, and the task quantity increases with the growth of environment size~\cite{zhu2021path} and environment numbers. Therefore, it is inefficient to improve the positioning accuracy by collecting more data. This fact motivates us to reduce the positioning error by optimizing the network architecture, given the constraint of dataset size. We get inspiration from recent network architectures proposed for image-related tasks and propose a highly-accurate positioning network.

\subsection{Tracking Systems w/o IMUs}
The data collection task for tracking systems w/o IMUs is much heavier than positioning systems~\cite{xie2020learning}. Existing data-driven CSI-based tracking systems usually integrate the CSI environments into the design process, making themselves environment-specific, and further increasing the data collection burden. This inspires us to design a universal tracking system independent from CSI environments. Motivated by~\cite{ilyas2016drift}, where the authors mentioned some environmentally-invariant motion patterns, we first introduce the idea of trajectory prior, and then adopt PnP to construct our tracking system. The proposed tracking system is compatible with arbitrary environment and IMUs with arbitrary precision.
%\subsection{Assumptions}
%Here, we clarify some important assumptions used in our work. First of all, before putting a positioning network into use, the parameters of the network should be trained on a CSI-location dataset. To construct the dataset, CSI is measured at some random RPs in the indoor environment with known locations, as shown in Fig~\ref{fig:system}. Also, as the CSI is environment-specific, the data-collection and training process is repeated for each indoor environment, and the positioning network deployed for each environment is different. Secondly, based on the fact that a MT will move from one indoor environment to another, the tracking system should be able to work across the environments. In addition, considering the time-consuming data-collection process for training a tracking system, it would be better that a tracking system trained under one environment can be applied to any other environment directly without retraining. Thirdly, since it is hard to ensure the IMU deployed on each MT has the same precision, the IMU-aided tracking system should be compatible with IMU with arbitrary precision.    

\section{CSI-fingerprinting Positioning Using Deep CNN}
\label{position}
A large RF means pixels far away from each other in the spatial and frequency dimensions of CSI can be considered simultaneously by the network~\cite{luo2016understanding}, which can be pretty helpful in exploiting location-related details. For example, it has been investigated in~\cite{tong2020csi} that using CSI phase differences from antennas that have larger spatial distances for angle-of-arrival (AoA) estimation has a better anti-noise performance. From~\cite{xiao2012fifs}, CSI amplitudes from the subcarriers whose distances are larger than the coherent bandwidth are more effective as fingerprints as these subcarriers are fading independently. Therefore, instead of focusing on the local pixels, we should collect and analyze the clues provided by non-neighbouring pixels.

However, two main factors restrict the RF of the existing CNN-based methods for CSI-fingerprinting indoor positioning: 1) 
%related works prefer to use convolutional
%layers with kernel shape $(1,n)$ for feature extraction~\cite{cerar2021improving,chin2020intelligent,de2020csi}, rather than typical $(n,n)$ kernels used in images-related tasks. This means these works only increase the receptive field along the frequency domain and ignores the channel responses in adjacent antenna elements, while phase difference of neighbouring antennas is highly-correlated to the angle-of-arrival (AOA) of the received signals, providing an important directional information of the target location~\cite{tong2020csi}, thus should be considered; Second, 
the number of the convolutional layers of the existing methods is not large enough~\cite{sun2019fingerprint,cerar2021improving,chin2020intelligent,
wu2021learning}, while the number of convolutional layers should be increased to enlarge the RF~\cite{kim2016accurate}. %Also, some works prefer to increase the number of channels of feature maps every time convolutional layers are applied, making the network parameters grow in an exponential way along with the growth of layers, which further prevents the CNN to go deeper.
2) the RF of a convolutional layer is naturally restricted by its inertial working mechanism, where each pixel in the output features of a convolutional layer is a weighted sum of the pixels in a small local region in the input features, making CNN itself inefficient in extracting the global context from CSI. %Third, some works utilizes convolutional layers with kernel shape $(1,n)$ for feature extraction~\cite{cerar2021improving,chin2020intelligent,de2020csi},

%To deal with the problems caused by the limited receptive field of CNN, some works tend to add multiple fully-connected (FC) layers with a large number of hidden units behind the CNN part to fuse the global information of the input~\cite{cerar2021improving,
%chin2020intelligent,arnold2019novel}. However, we assume it would be better to increase the receptive field of the CNN part rather than relying on the FC layers for global context as it has been proved CNN is more effective in feature extraction process compared with FCNN~\cite{chin2020intelligent}. 

Different from the existing works, we propose two ways to increase the RF of our network. On the one hand, we build a very deep CNN with more than 20 convolutional layers by stacking residual blocks (RBs). On the other hand, we use an attention-augmented residual block (AARB) for global context extraction, with which it is easier for the RF to fully fill the whole input. More details about our positioning network are given hereafter.
\begin{figure}
\centering
\includegraphics[scale=0.33]{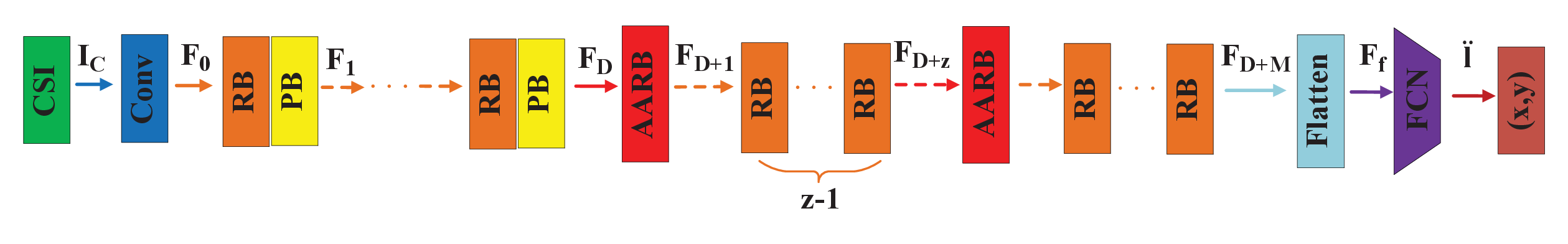}
\caption{The overall architecture of the proposed attention-augmented residual convolutional neural network (AAResCNN).}
\label{fig:arc}
\end{figure}
\subsection{Network Architecture}
\label{archi}
We first present the overall network architecture of the proposed AAResCNN for CSI-based fingerprinting indoor localization. As shown in Fig~\ref{fig:arc}, our network mainly consists of four parts: residual blocks (RBs), pooling blocks (PBs), attention-augmented residual blocks (AARBs), and finally a FCN. We denote the input CSI and output predicted location as $I_{C}$ and $\ddot{l}$ respectively. First of all, we use one convolutional layer to extract shallow features $F_{0}$ from $I_{C}$,
\begin{equation}
F_{0}=Conv(I_{C}),
\end{equation}
where $Conv(\cdot)$ denotes convolution operation. Considering the large spatial size ($H\times W$) of CSI in MIMO-OFDM systems, we then use a cascade of RBs and PBs to fuse the local information in shallow features $F_{0}$ and reduce its spatial size gradually. Assuming we have D cascaded RBs and PBs in total, the output of the $d$-th PB can be obtained as
\begin{equation}
\begin{aligned}
F_{d}&=PB_{d}(RB_{d}(F_{d-1}))\\
&=PB_{d}(RB_{d}(\cdots PB_{1}(RB_{1}(F_{0}))\cdots)),
\end{aligned}
\end{equation}
where $PB_{d}$ and $RB_{d}$ denote the $d$-th PB and RB, respectively, and the value of $D$ is set according to the size of CSI input. If the pooling size and stride are set as $(p,q)$ in PBs, the height and width of $F_{D}$ will be $1/(pD)$ and $1/(qD)$ of those in $F_{0}$. More details about RB and PB will be given in Section~\ref{sec rb} and Section~\ref{sec pb}, respectively.

After extracting spatially-downsampled features $F_{D}$, we continue to adopt a stack of RBs to extract local information from $F_{D}$. Also, to better capture the global context, we substitute RB with AARB after every $z$ RBs. If $z=1$, we only use AARB after $F_{D}$. If there are $M$ blocks after the last PB, the $(D+m)$-th output feature maps of our network can be expressed as
\begin{equation}
F_{D+m}=\begin{cases}
AARB(F_{D+m-1}), & mod(m-1,z)=0,\\
RB(F_{D+m-1}), & mod(m-1,z)\neq 0,
\end{cases}
\end{equation}
where $mod(\cdot)$ denotes the modulo operation. More details about AARB will be shown in Section~\ref{sec aarb}. After applying $M$ blocks to $F_{D}$, we flatten the deep features $F_{D+M}$ to a vector $F_{f}$,
\begin{equation}
F_{f}=Flatten(A_{M}),
\end{equation}
%where $Flatten(\cdot)$ denotes the flatten function.

Finally, we use a 3-layer FCN with a configuration of $n_{1}$-$n_{2}$-2 hidden neurons to gradually map $F_{f}$ from the high-dimensional feature space to 2D coordinates. The predicted location can be obtained by
\begin{equation}
\ddot{l}=FCN(F_{f}).
\end{equation}

Supposing there are $I$ CSI-location pairs in the training dataset in total, the overall network is trained to minimize
the following loss function,
\begin{equation}
\text{Loss}= \sum_{i=1}^{I}\|\text{Position-Net}(I_{C}^{i})-l^{i}\|^2=\sum_{i=1}^{I}\|\ddot{l}^{i}-l^{i}\|^2,
\end{equation}
where $\|\cdot\|^2$ denotes $\textit{l}_{2}$-norm and $l^{i}$ is the true location of the $i$th training sample. $I_{C}^{i}$ is the CSI measured at location $l^{i}$ and $\ddot{l}^{i}$ is the estimated location using $I_{C}^{i}$ as input.
%where $FCN(\cdot)$ denotes the 3-layer fully-connected network with activation layer inserted in between. 
\subsection{Going deeper with residual blocks}
\label{sec rb}
\begin{figure}[t]
\centering
\subfloat[\centering Residual block (RB)]{
\includegraphics[width=0.28\textwidth]{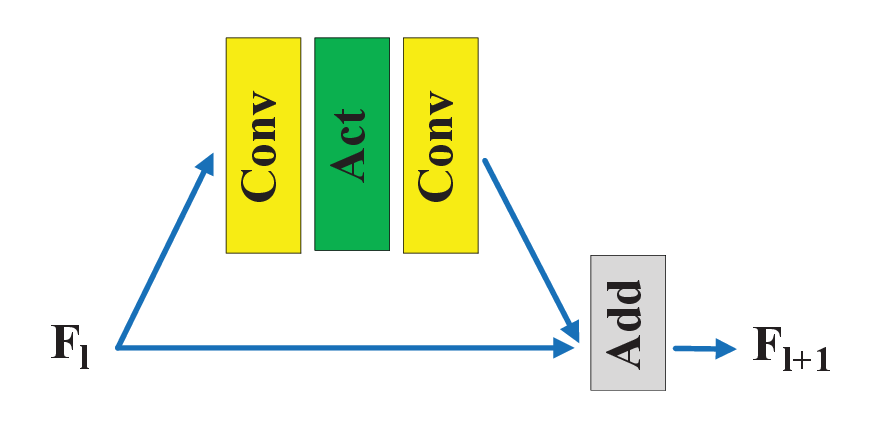}
}
\quad
\subfloat[\centering Attention-augmented residual block (AARB)]{
\includegraphics[width=0.3\textwidth]{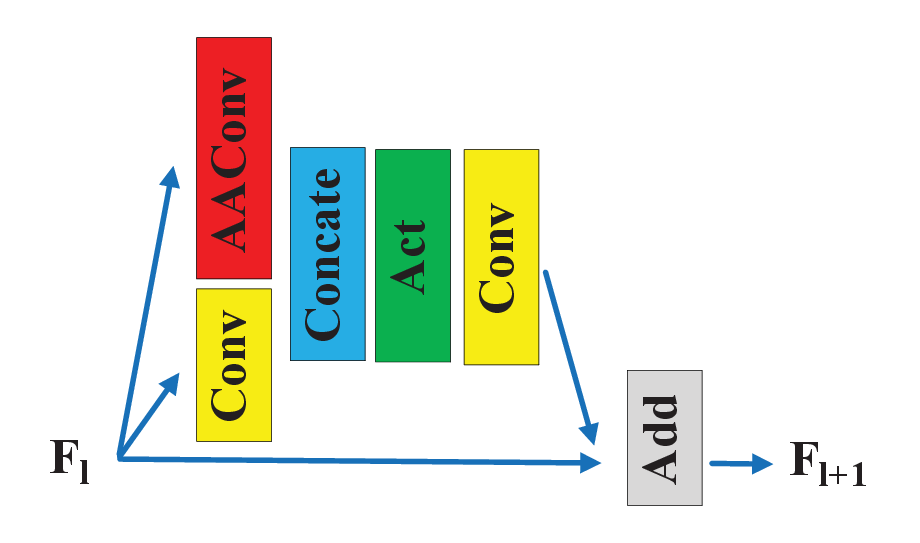}
}
\\
\caption{The architecture of residual block (RB) and attention-augmented residual block (AARB).}
\label{fig:res}
\end{figure}

To create filters that are responsive to a larger region in the input, increasing the depth of networks has been widely used in image-related tasks~\cite{kim2016accurate}. For example, if we use filters of size $k\times k$ for all the convolutional layers in our network, the RF of the $d$-th convolutional layers in our network will be size of $((k-1)d+1)\times ((k-1)d+1)$, which grows as $d$ increases. Another benefit of using a very deep network is that it can introduce more nonlinearities to the network as a convolutional layer is often followed by a non-linear layer. With the growth of nonlinearities, our network can model more complex functions, which is essential when dealing with CSI in complicated wireless environments. 

However, if we stack convolutional layers in a plain way, we may face a performance degradation problem. That is, as the network goes deeper, the newly-added layers not only fail to bring performance improvements, but also raise the training and test errors, which is also reported in~\cite{he2016deep}.
To address the degradation problem, we choose to adopt the residual blocks (RBs) proposed in~\cite{lim2017enhanced}. We show the detailed architecture of one RB in Fig~\ref{fig:res}(a). As we can see from Fig~\ref{fig:res}(a), one residual block consists of two convolution layers, an activation layer, along with an identity skip connection. The skip connection provides convenience for the back-propagation of gradients, which helps to stabilize the training phase and reduce the training pressure. Therefore, by stacking RBs, we can build a very deep CNN with a large RF.

Note that in our block-based CNN architecture, the channel dimension of feature maps keeps constant across different layers, 
%which means the number of parameters is growing in a linear way with the growth of layers in our network, 
different from the channel growth strategies in~\cite{chin2020intelligent,wu2021learning}. Actually, keeping the channel number a small constant can lead to a lighter model. For example, the model parameters used to build a convolutional layer between feature maps with 256 channels can build up to 64 convolutional layers among features with 32 channels.

\subsection{Pooling blocks}
\label{sec pb}
\begin{figure}[t]
\centering
\includegraphics[scale=0.3]{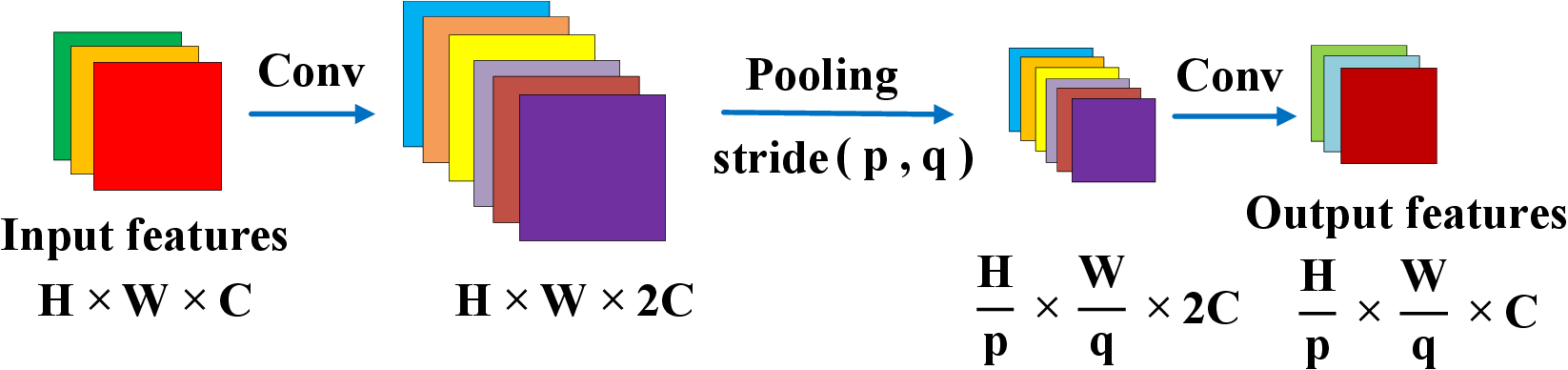}
\caption{The architecture of pooling blocks (PBs).}
\label{fig:pb}
\end{figure}
Average pooling (AveP) is widely used in CSI-based fingerprinting indoor positioning networks to reduce the feature size. After an AveP layer with equal pooling size and stride $(p,q)$, the height and width of input features will be reduced to $1/p$ and $1/q$ of the original sizes while the channel number will keep unchanged. However, information loss is unavoidable in this process since pixels in the height and width dimensions are averaged manually and represented by fewer pixels. To mitigate the information loss during the pooling process, we introduce the idea of pooling blocks (PBs)~\cite{lin2017refinenet}. In a PB, we first use a convolutional layer to double the channel number of input features. Next, we apply an AveP to reduce the spatial size. Finally, we adopt another convolutional layer to reduce the channel number to the original size. It is expected the added two convolutional layers will learn to transfer the important information for position estimation from spatial dimension to channel dimension and preserve it through training. The details of a PB are shown in Fig~\ref{fig:pb}. The idea of doubling the channel number before downsampling is also used in~\cite{hu2019runet}.
\subsection{Attention-augmented Residual blocks}
\label{sec aarb}
Although the theoretical receptive field grows as more RBs are stacked, the growth rate of effective receptive field (ERF) is much slower~\cite{luo2016understanding}.  Also,  as analyzed in~\cite{luo2016understanding}, not all pixels in the RF contribute equally to the outputs. In fact, ERF follows Gaussian distribution, which means central pixels in the RF usually have larger impact\textcolor{blue}{s}. To overcome these drawbacks, attention-mechanisms have been widely adopted~\cite{Lin_Li_Zheng_Cheng_Yuan_2020}.  

Attention mechanisms have recently been proposed for global context extraction in language models~\cite{vaswani2017attention} and image-related tasks~\cite{dosovitskiy2020image}. By introducing attention mechanisms, neural networks can see the whole input rather than focusing on a local region. In this work, we explore the performance of attention-augmented convolution (AAConv) proposed in~\cite{bello2019attention} for CSI-fingerprinting indoor localization, and use attention-augmented residual blocks (AARBs) to enhance the performance of positioning networks. %Note that besides AAConv, vision transformer~\cite{dosovitskiy2020image,chen2021pre} is another popular approach to introduce attention mechanism. However, vision transformer is based on RNN architecture and incompatible with the concept of RBs in this work, therefore, we will be evaluated in our future work. 
Hereafter, we give a detailed description of AAConv and AARBs. 
% (c) the detials of attention-augmented convolutional layer with one head.
\begin{figure}[t]
\centering
\subfloat[\centering Traditional convolution layer (Conv)]{
\includegraphics[width=0.26\textwidth]{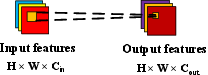}
}
\quad\quad
\subfloat[\centering Attention-augmented convolution layer (AAConv)]{
\includegraphics[width=0.3\textwidth]{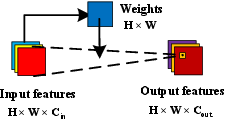}
}
\\
\caption{The comparison of traditional convolution layer (Conv) and attention-augmented convolution layer (AAConv).}
\label{fig:aa-ab}
\end{figure}
\subsubsection{Comparison of Conv and AAConv}
%{\bf (a) Comparison of Conv and AAConv.} 
We first compare the overall architectures of the traditional convolution layer (Conv) and the AAConv in Fig~\ref{fig:aa-ab}. Specifically, in the Conv, each pixel in the output features is a weighted sum of pixels in a local window in the input features and the weights are shared for all output pixels, while all pixels in the input features are considered in the AAConv, and the weights are dynamically generated for each output pixel. By introducing the AAConv, the RF increases from a local window to the whole input, which significantly improves the feature extraction efficiency for the global context. 
\subsubsection{Attention-augmented convolutional layer with a single head}
\label{aarb-1}
% (c) the detials of attention-augmented convolutional layer with one head.
\begin{figure}[t]
\begin{center}
\includegraphics[scale=1.1]{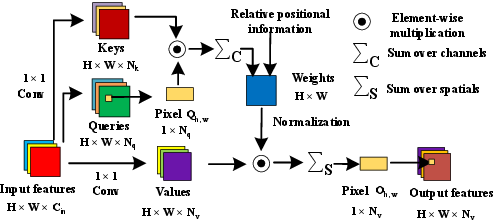}
\caption{The architecture of attention-augmented convolutional layer with a single head}
\label{fig:aa-c}
\end{center}
\end{figure}
We now describe the detailed architecture of the AAConv with a single head, which is also depicted in Fig~\ref{fig:aa-c}. Given input features with shape $(H,W,C_{in})$, where $H$, $W$, and $C_{in}$ denotes the height, width and channels of the input feature, respectively, the AAConv first computes three intermediate features called queries $Q\in \mathcal{R}^{H\times W\times N_{q}}$, keys $K\in \mathcal{R}^{H\times W\times N_{k}}$ and values $V\in \mathcal{R}^{H\times W\times N_{v}}$ through 3 sets of $1\times 1$ Conv, where $N_{q}$ equals to $N_{k}$ all the time. 

Next, for each pixel $(h,w)$ in the height and width dimensions of the output features $O\in \mathcal{R}^{H\times W\times N_{v}}$, which is also denoted as $O_{h,w}\in \mathcal{R}^{N_{v}}$, an attention weight matrix $w^{O_{h,w}}\in \mathcal{R}^{H\times W}$ is computed from queries $Q$ and keys $K$ through
\begin{equation}
w^{O_{h,w}}_{m,n}=\frac{Q_{h,w}\cdot K_{m,n}}{\sqrt{N^{q}}} \quad \text{for}\, m=1,\ldots,H; \,n=1,\ldots,W
\label{equ:2}
\end{equation}
where $w^{O_{h,w}}_{m,n}\in \mathcal{R}$ denotes the pixel $(m,n)$ in the height and width dimensions of attention weight $w^{O_{hw}}$, $Q_{h,w}\in \mathcal{R}^{N_{q}}$ denotes the pixel $(h,w)$ in queries $Q$, and $K_{m,n}\in \mathcal{R}^{N_{k}}$ denotes the pixel $(m,n)$ in keys $K$; $\cdot$ denotes inner product. 

The values in the attention weight matrix, $w^{O_{h,w}}$, denote the importance of each pixel in keys $K$ to $Q_{h,w}$ by analysing the feature space, and the weight matrix is then normalized by
\begin{equation}
\hat{w}^{O_{h,w}}=\text{Softmax}_{2D}(w^{O_{h,w}}),
\label{equ:3}
\end{equation}
where $\text{Softmax}_{2D}(\cdot)$ denotes applying $\text{Softmax}$ function to a matrix. Specifically, 
for a matrix $w\in \mathcal{R}^{H\times W}$, $
[\text{Softmax}_{2D}(w)]_{h,w}=e^{w_{h,w}}/\sum_{m=1}^{H}\sum_{n=1}^W e^{w_{m,n}}\,\, \text{for}\, h=1,\ldots,H; \,w=1,\ldots,W
$, where $w_{h,w}$ denote pixel $(h,w)$ in $w$.

Finally, the normalized attention weight matrix, $\hat{w}^{O_{h,w}}$, is used to reweigh values $V$. The reweighing scheme can be formulated as,
 \begin{equation}
O_{h,w}=\sum_{m=1}^{H}\sum_{n=1}^{W}\hat{w}^{O_{h,w}}_{m,n}V_{m,n}
\label{equ:4}
\end{equation}
where $\hat{w}^{O_{h,w}}_{m,n}\in \mathcal{R}$ denotes pixel $(m,n)$ in weight matrix $\hat{w}^{O_{h,w}}$, $V_{m,n}\in \mathcal{R}^{N_{v}}$ denotes pixel $(m,n)$ in values $V$. 

From~(\ref{equ:2}),~(\ref{equ:3}) and~(\ref{equ:4}), all pixels in the height and width dimensions of input features are considered and the pixels that are more important to pixel $(h,w)$ in the feature space will contribute more to the generation of $O_{h,w}$.
\iffalse
\begin{figure}[t]
\centering
\includegraphics[scale=1.3]{images/attention_head.eps}
\caption{The architecture of attention augmented convolutional layer with $N_{h}$ heads.}
\label{fig:aa-head}
\end{figure}
\fi

In the original AAConv proposed in~\cite{bello2019attention}, the AAConv with multiple heads is utilized to explore the global context in a more diversified way rather than a single head. However, we notice that using multi-heads will significantly increase the computational cost. Therefore, we only use AAConv with a single head here.
\iffalse
\subsubsection{Attention-augmented convolutional layer with multiple heads}
\label{aarb-head}
From a set of queries $Q$, keys $K$, and values $V$, we can get a single result $O$. While in practice, we usually run multiple groups of queries $Q$, keys $K$, and values $V$ in parallel for each AAConv layer to explore the global context in a more diversified way. Specifically, to realize AAConv with $N_{h}$ heads, the $N_{h}$ groups of attention-augmented results are first concatenated along the channel dimension, and then processed by a $1\times 1$ Conv, %which is used to fuse the output features under different attention mechanisms, 
as illustrated in Fig~\ref{fig:aa-head}.
\fi
\subsubsection{Attention with relative positional information}
The last thing about the AAConv is the relative positional information, without which AAConv is permutation equivariant~\cite{bello2019attention}. Permutation equivariant means that for any permutation $\pi$ of the pixel locations, the equation $AAConv(\pi(X))= \pi(AAConv(X))$ holds for AAConv, where $X$ denotes the input features. Permutation equivariant is an undesirable property for CSI-fingerprinting localization as it means the structural information in CSI is not utilized effectively.
%Although global contexts are considered for attention mechanisms in Section~\ref{aarb-1} and Section~\ref{aarb-head}, the resulting output features $O$ fails to preserve the structural information in input features. Denoting input features as $X$, the output $O(X)$ is actually permutation equivariant, which means $O(\pi(X))=\pi(O(X))$ holds for any permutation $\pi(\cdot)$ of the pixel locations in $X$. 
To encourage the attention mechanism to consider both the pixel locations and feature similarity, I. Bello et al. in~\cite{bello2019attention} incorporate relative positional encodings~\cite{shaw2018self} to~(\ref{equ:2}):
\begin{equation}
\begin{aligned}
w^{O_{h,w}}_{m,n}=&(\frac{1}{\sqrt{N_{q}}})(Q_{h,w}\cdot K_{m,n}+Q_{h,w}\cdot r_{m-h}^{H}\\
&+Q_{h,w}\cdot r_{n-w}^{W})\quad \text{for}\, m=1,\ldots,H;\, n=1,\ldots,W
\end{aligned}
\label{equ:5}
\end{equation}
%\quad 
where $m-h$ and $n-w$ denote the relative height distance and width distance between pixel $(h,w)$ and $(m,n)$, respectively, and $r_{m-h}^{H}\in \mathcal{R}^{N_{q}}$ and $r_{n-w}^{W}\in \mathcal{R}^{N_{q}}$ are learnt positional encodings for relative height $m-h$ and width $n-w$, respectively. For an input feature of spatial size $(H,W)$, $m-h$ takes the values from $-(H-1)$ to $(H-1)$ and $n-w$ from $-(W-1)$ to $(W-1)$, therefore, the total number of positional encodings for one layer is $2(H+W)-2$.

\subsubsection{Attention-augmented residual blocks}
After introducing the AAConv, we now give the architecture of the proposed AARB, which is shown in Fig~\ref{fig:res}(b). As the Conv captures local information while the AAConv extracts global context, we combine the Conv and the AAConv following the idea in~\cite{bello2019attention}. Specifically, we replace the first Conv layer in a RB (see Fig~\ref{fig:res}(a)) with a Conv and an AAConv, and their outputs are concatenated along the channel dimension, which results in an increased number of channels $N_{v}+C_{in}$, while the second Conv layer in an AARB will fuse the information in concatenated features and reduce the channel number to $C_{in}$.
\subsection{Complexity Analysis}
We compare the number of parameters and computational complexity of three basic operations in our network, \textit{i}.\textit{e}., AveP, Conv, AAConv. Following~\cite{bello2019attention}, we use floating point operations (FLOPs) as the index of computational cost, which is defined as the number of addition and multiplication used in one operation. Supposing the input features ha\textcolor{blue}{ve} shapes of $H\times W\times C_{in}$, the results are shown in Table~\ref{table0}. As $N_{q}$ and $N_{v}$ are usually set as small values, the number of parameters of an AAConv layer is usually smaller than a Conv layer. However, the computational cost of an AAConv layer $(\textit{O}((HW)^2N_{q}))$, which is an exponential function of spatial size $HW$, is larger than that of a Conv layer $(\textit{O}(HWk^2C_{in}^2))$, especially when spatial size is large. Therefore, we only use AARB on spatially-downsampled features.
\begin{table}[t]
\caption{\label{table0} The Number of Parameters and FLOPs of different operations}
\begin{center}
\resizebox{0.55\columnwidth}{!}{\begin{tabular}{|c|c|c|c|}
\hline
Operation & AveP         & Conv   & AAConv \\ \hline
Params Num     & 0 & $\approx k^{2}C_{in}^{2}$ & $\approx C_{in}(2N_{q}+N_{v})+2(H+W)$  \\ \hline
Flops     & $HWC_{in}$ & $\approx 2k^{2}HWC_{in}^{2}$ & \begin{tabular}[c]{@{}c@{}}$\approx 2HW[HW(3N_{q}+N_{v})$\\$+C_{in}(2N_{q}+N_{v})]$ \end{tabular} \\ \hline

\end{tabular}}
\end{center}
\end{table}
\section{Tracking System via Deep Trajectory Prior}
\label{tracking}
In this section, we propose a universal tracking system by exploring the properties of trajectories based on the assumption that some motion patterns are environmentally-invariant. %which is designed separately from the CSI-based positioning system.
\subsection{Trajectory Refinement with Deep
Prior}
Let $\ddot{L}\in \mathcal{R}^{T\times 2}$ denote the predicted positions along a trajectory from a positioning system in consecutive $T$ timestamps. Let $L\in \mathcal{R}^{T\times 2}$ denote the true positions of the same trajectory. The relationship between $\ddot{L}$ and $L$ is
\begin{equation}
\ddot{L}=L+v_{p}
\label{equ:t1}
\end{equation}
where $v_{p}$ is the measurement noise of the positioning system, which is assumed to be additive white Gaussian noise with zero mean\footnote{To show the effectiveness of the Gaussian noise assumption, we plot the positioning error distributions of the trained AAResCNNs in the considered test datasets in Section~\ref{Error-distribution}}. %and standard deviation $\sigma_{v_{p}}$ for simplicity. %although the noise distribution assumption may be different from the real one. %we find it works quite well when the tracking system is applied to a real positioning system. 
The purpose of trajectory refinement is to recover the clean trajectory $L$ from its noisy measurement $\ddot{L}$. Since this is an ill-posed inverse problem, a regularization term, which is also called a prior function, is required to constrain the solution space. %From a Bayesian perspective, Equation~\ref{equ:t1} equals to solve a maximum a posteriori (MAP) problem,
%\begin{equation}
%\hat{L}=\mathop{\arg\max}\limits_{L}\quad \log p(\tilde{L}|L)+\log p(L)
%\label{equ:t2}
%\end{equation}
%where $p(\tilde{L}|L)$ denotes the likelihood of measurement $\tilde{L}$ given $L$, $p(L)$ denotes the prior of $L$ , which is independent of the positioning network, and $\hat{L}$ denotes the refined trajectory from $\tilde{L}$. 
Therefore, the trajectory refinement problem can be reformulated as
\begin{equation}
\hat{L}=\mathop{\arg\min}\limits_{L}\quad \frac{1}{2}\|\ddot{L}-L\|^2+\lambda \phi(L),
\label{equ:t3}
\end{equation}
where $\|\ddot{L}-L\|^2$ denotes the fidelity term that ensures that refined trajectory $\hat{L}$ should be close to the measurement $\ddot{L}$; $\phi(L)$ denotes the trajectory prior function, which enforces the refined trajectory $\hat{L}$ should have some desired properties that a true trajectory $L$ should have; and finally, $\lambda$ denotes the trade-off parameter that balances the effect of fidelity term and the prior term in trajectory refinement, whose value should be tuned during experiments. 

According to Bayesian probability,~(\ref{equ:t3}) corresponds to denoising $\ddot{L}$ when it has been corrupted by additive Gaussian noise of standard deviation $\sqrt{\lambda}$ with a Gaussian Denoiser, where the trajectory prior $\phi(x)$ is implicitly replaced by a denoiser prior~\cite{zhang2017learning}\cite{venkatakrishnan2013plug}. To address this, we rewrite~(\ref{equ:t3}) as
\begin{equation}
\hat{L}=Denoiser(\ddot{L},\sqrt{\lambda})
\label{equ:t4}
\end{equation}
%\textit{i}.\textit{e}.

There are two different ways to solve~(\ref{equ:t4}), \textit{i}.\textit{e}., a model-based optimization method and a discriminative learning method~\cite{zhang2017learning}. In this paper, instead of using a hand-crafted prior, we propose to learn the prior with deep learning based methods.
To this end, we assume that a trajectory dataset exists, which
contains all kinds of trajectories in indoor environments.%Instead of using a hand-crafted prior, we assume a trajectory dataset that contains all kinds of indoor trajectories exits and propose to learn the prior from the trajectory dataset.
\subsection{Denoising Network for Tracking}
\label{sec:track process}
\begin{figure}[t]
\centering
\subfloat[Training phase]{
\includegraphics[width=0.23\textwidth]{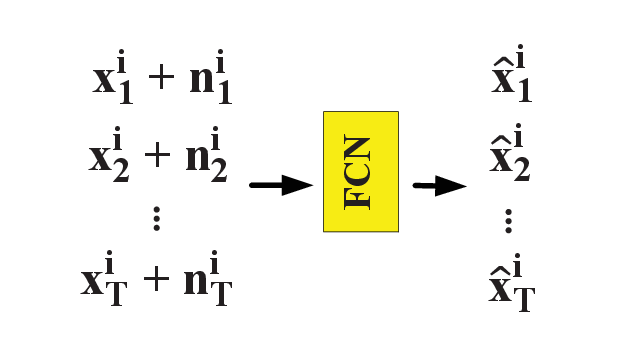}
}
\quad
\subfloat[Testing phase]{
\includegraphics[width=0.35\textwidth]{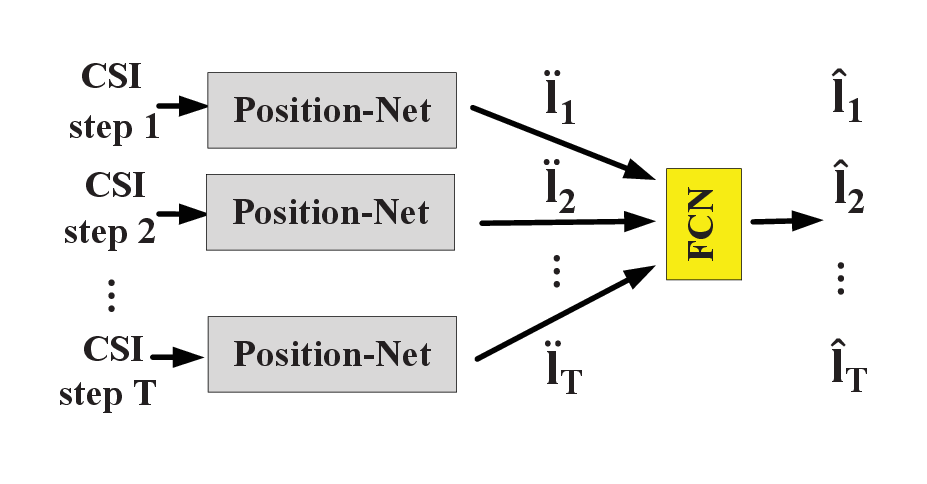}
}
\\
\caption{The architecture of the tracking system during the training phase and the testing phase.}
\label{fig:traintest}
\end{figure}

%In the context of indoor tracking, the current location of a moving target, \textit{i}.\textit{e}., $x_{T}$, depend not only on the current measurement $y_{T}$, but also on the previous measurements $y_{1},\ldots,y_{T-1}$, which is quite similar to the idea of RNN-based neural network where the current output is determined by both the current input and the historical data. Therefore, it is straightforward to adopt a RNN-based model for the denoising problem depicted in Equation~\ref{equ:t4}. There are a variety of RNN models, such as vanilla RNN, LSTM~\cite{hochreiter1997long} and GRU~\cite{cho2014properties}. Among them, we chose LSTM to capture the trajectory prior from the noisy measurement due to its popularity in dealing with temporal sequence data. 
%$n_{i}\sim \mathcal{N}(0,\,(\sqrt{\lambda})^2)$

Suppose there are $N_{L}$ trajectories in the collected dataset $X_{L}=\{x^{i}\}_{i=1}^{N_{L}}$, where $x^{i}\in \mathcal{R}^{T\times 2}$ denotes one trajectory sample that lasts $T$ timestamps. According to~(\ref{equ:t4}), as our aim is to learn a Gaussian denoiser based on a data-driven trajectory prior, we first manually add random \textit{i}.\textit{i}.\textit{d}. Gaussian noise with zero mean and standard deviation equal to $\sqrt{\lambda}$ to trajectory $x^{i}$,
\begin{equation}
y^{i}_{t}=x^{i}_{t}+n^{i}_{t} \quad t=1,\ldots,T 
\label{equ:t5}
\end{equation}
where $y^{i}_{t}$, $x^{i}_{t}$, and $n^{i}_{t}$ denote the noisy position, the true position, and the random noise on trajectory $x^{i}$ at time $t$, respectively. Next, we flatten $y^{i}$ to be a vector $y^{i}_{f}\in \mathcal{R}^{2T}$ and adopt a 3-layer FCN with 128 hidden units to refine $y^{i}_{f}$,
\begin{equation}
\hat{x}^{i}_{f}=W_{3}\mathit{f}(W_{2}\mathit{f}(W_{1}y^{i}_{f}+b_{1})+b_{2})+b_{3},
\label{equ:t6}
\end{equation}
where $\hat{x}^{i}_{f}\in \mathcal{R}^{2T}$ denotes the flatten version of the denoised trajectory, $W_{1}$, $W_{2}$, $W_{3}$ denote the weights in FCN and $b_{1}$, $b_{2}$, $b_{3}$ denote the biases, respectively. $\mathit{f}(\cdot)$ denotes the activation layer. Then, we reshape $\hat{x}^{i}_{f}$ to $\hat{x}^{i}\in \mathcal{R}^{T\times 2}$. Finally, the FCN is trained by,
\begin{equation}
\min \frac{1}{N_{L}T}\sum_{i=1}^{N_{L}}\sum_{t=1}^{T} \|\hat{x}^{i}_{t}-x^{i}_{t}\|^2.
\label{equ:t8}
\end{equation}

After training, we can get a Gaussian Denoiser under a specific value of $\lambda$. As the optimal value of $\lambda$ is unknown before deployment, we train a series of Gaussian Denoisers under different values of $\lambda$ in advance so that we can choose the one with the best performance in the testing phase. During the online testing phase, we take trajectory $\ddot{L}$ recovered by a positioning system as inputs and refine it through the deep tracking network, as shown in~(\ref{equ:track}). For positioning systems
with varying accuracy, we only need to fine-tune the value
of $\lambda$. We show the training phase and testing phase of the tracking system in Fig~\ref{fig:traintest}.

%Note that the best $\lambda$ depends on the accuracy of the positioning network. Once the positioning network is deployed, the best Gaussian Denoiser is also fixed.
%As CSI information and CSI-based positioning network are not necessary during the training phase, our denoising network-based tracking system can be applied to any wireless indoor environment. The deployment cost consists of collecting a trajectory dataset and the aforementioned training process. The performance gain of our tracking system solely comes from inertial properties of the universal motion patterns in indoor environments.    

\section{Plug and Play for IMU measurements}
\label{imu tracking}
%To use sensory data, a straightforward way is to take it as an extra input of the tracking network alongside the measurements from the positioning system. However, there are a variety of IMU equipments in reality and their accuracy on the step length and moving head estimation is different~\cite{yan2019performance}. In other words, the measurement noise of different IMU equipments is diverse. Although we can train a series of tracking network in parallel under all kinds of noise level, it is a large burden for the training process. Also, storing these different tracking networks is memory-inefficient. Thus, one network for all IMUs is on demand.
In this section, we adopt the idea of PnP~\cite{venkatakrishnan2013plug,zhang2017learning} to incorporate the IMU measurements with different precisions into an existing tracking system. 
%$M=[m_{2},m_{3},\cdots,m_{T}] \in \mathcal{R}^{T-1}$ denote the IMU measurements in consecutive $T$ timestamps, where 

Let $\tilde{m}_{t}=[\tilde{r}_{t},\tilde{\theta}_{t}]$ denote the estimated walking distance and direction from time $t-1$ to time $t$ by the IMU. Suppose the true step length and direction in this period are $r_{t}$ and $\theta_{t}$, respectively. We assume the following relationship exists,
\begin{equation}
\tilde{r}_{t}\cos(\tilde{\theta}_{t})=r_{t}\cos(\theta_{t})+v_{x},\,
\tilde{r}_{t}\sin(\tilde{\theta}_{t})=r_{t}\sin(\theta_{t})+v_{y}
\label{equ:t18}
\end{equation}
where $v_{x}\sim \mathcal{N}(0,\sigma_{v_{x}})$ and $v_{y}\sim \mathcal{N}(0,\sigma_{v_{y}})$ denote the measurement noise of the IMU on $x$ and $y$ coordinates, respectively. We also define the SNR of IMU measurements as,
\begin{equation}
SNR_{IMU}=20\log\frac{|r_{t}\cos(\theta_{t})|}{\sigma_{v_{x}}}=20\log\frac{|r_{t}\sin(\theta_{t})|}{\sigma_{v_{y}}}
\label{equ:t19}
\end{equation}

Let $\tilde{M}=[\tilde{m}_{2},\tilde{m}_{3},\cdots,\tilde{m}_{T}] \in \mathcal{R}^{T-1}$ denote the IMU measurements in consecutive $T$ timestamps. After considering IMU measurements, the trajectory refinement problem becomes
\begin{equation}
\bar{L}=\mathop{\arg\min}\limits_{L}\quad f(L,\tilde{M})+\mu \|\ddot{L}-L\|^2+\hat{\lambda} \phi(L),
\label{equ:t9}
\end{equation} 
where
\begin{equation}
f(L,\tilde{M})=\sum_{t=1}^{T-1}\|(l_{t}-l_{t-1})-(\tilde{r_{t}}\cos(\tilde{\theta_{t}}),\tilde{r_{t}}\sin(\tilde{\theta_{t}}))\|^2,
\label{equ:t17}
\end{equation} 
and $f(L,\tilde{M})$ enforces refined trajectory $\bar{L}$ to satisfy the motion constraints from IMU measurements; $\|\ddot{L}-L\|^2$ encourages refined trajectory $\bar{L}$ to be close to estimated trajectory $\ddot{L}$ from the positioning network; $\phi(L)$ denotes refined trajectory $\bar{L}$ should match the deep trajectory prior. Furthermore, $\mu$ and $\hat{\lambda}$ are the trade-off parameters used to balance the contribution of IMU measurements ($f(L,\tilde{M})$), positioning network ($\|\ddot{L}-L\|^2$), and trajectory prior ($\phi(L)$) in the tracking process. In general, for IMU measurements under different values of SNR, we can fine-tune the values of $\mu$ and $\hat{\lambda}$ to incorporate them correctly.

To solve~(\ref{equ:t9}), a new variable $Z$ is introduced to decouple the model-based parts and network-based parts,
\begin{equation}
\begin{split}
&\mathop{\min}\limits_{L,Z}\quad f(L,\tilde{M})+\mu \|\ddot{L}-L\|^2+\hat{\lambda} \phi(Z)\\
&\,\mathrm{s.t.} \quad \quad L=Z,
\end{split}
\label{equ:t10}
\end{equation} 
which can now be solved by the alternating direction method of multipliers (ADMM)~\cite{boyd2011distributed}. Specifically, the $k$-th iteration in solving~(\ref{equ:t10}) with the ADMM algorithm can be written as, 

\begin{equation}
L^{k+1}=\mathop{\arg\min}\limits_{L}\,f(L,\tilde{M})+\mu \|\ddot{L}-L\|^{2}+\rho/2\|L-Z^{k}+p^{k}\|,
\label{equ:t11}
\end{equation}

\begin{equation}
Z^{k+1}=\mathop{\arg\min}\limits_{Z} \quad \hat{\lambda} \phi(Z)+\rho/2\|L^{k+1}-Z+p^{k}\|,
\label{equ:t12}
\end{equation}

\begin{equation}
p^{k+1}=p^{k}+L^{k+1}-Z^{k+1},
\label{equ:t13}
\end{equation}
where $p$ is the scaled dual variable, $\rho$ is the penalty parameter, which does not affect the final result but controls the convergence rate of the ADMM algorithm. The updates of $L$ and $p$ can both be calculated mathematically. For the update of $Z$, following the same idea as in the previous section, we can reformulate~(\ref{equ:t12}) as 
\begin{equation}
Z^{k+1}=Denoiser(L^{k+1}+p^{k},\sqrt{\frac{\hat{\lambda}}{\rho}})
\label{equ:t16}
\end{equation}
which means~(\ref{equ:t12}) can be solved by using a Gaussian Denoiser trained under the noise level $\sqrt{\frac{\hat{\lambda}}{\rho}}$. As the value of $\rho$ has little influence on the final output, we can fix the value of $\sqrt{\frac{\hat{\lambda}}{\rho}}$ used to train the Gaussian Denoiser and fine-tune the value of $\rho$ instead when we fine-tune the value of $\hat{\lambda}$ to deal with IMU measurements with varing SNR. 
%In this way, we successfully construct an IMU-aided tracking system wh.

\section{Experiments}
\label{experiments}
In this section, we will first compare the performance of AAResCNN with SOTA methods~\cite{sun2019fingerprint,arnold2019novel,
chin2020intelligent,de2020csi,liu2022mhsa,wang2018deep,hoang2020cnn} on publicly-released CSI-based indoor positioning datasets. And then, we will verify the generality of our tracking system across CSI environments. Finally, we will show the flexibility of our IMU-aided tracking system in dealing with IMU measurements with varying precision.
\subsection{CSI-fingerprinting Positioning}
\subsubsection{Datasets}
To show the performance of our CSI-fingerprinting indoor positioning network, we conduct experiments on one publicly-released indoor LOS dataset: the KU Leuven Boardroom dataset\footnote{\url{https://homes.esat.kuleuven.be/~sdebast/measurements/measurements_boardroom.html}}, and one NLOS dataset: the KU Leuven Lab NLOS dataset\footnote{\url{https://homes.esat.kuleuven.be/~sdebast/measurements/measurements_lab.html}}~\cite{de2020mamimo,de2020csi}. The datasets contain CSI and position tags of a massive MIMO-OFDM system measured by the National Instruments 5G Massive MIMO testbed\footnote{https://www.ni.com/en-gb/innovations/white-papers/14/5g-massive-mimo-testbed--from-theory-to-reality--.html}. The system has $64$ antennas and $100$ sub-carriers, therefore, the measured CSI at each position has a shape of $(64,100,2)$. During the measurement, the users move along the predefined routes in $4$ grids, each spanning a $1.25$ m by $1.25$ m area. The CSI is collected at 5 mm intervals, resulting in $252004$ CSI samples in total. During the experiments, we do not use the total dataset for experiments as it is time-consuming and sometimes impractical to scan the indoor environment at mm level, especially for large shopping mall or conferencing room. Following the transfer learning experiment in~\cite{de2020csi}, we sample a smaller dataset. Specifically, we randomly select $5,000$ samples as the test dataset, $5,000$ sample as the validation dataset, and we change the number of the training samples, $I$, from $1,000$ to $5,000$ to $10,000$ to evaluate the influence of RPs' number on the positioning accuracy. Furthermore, as the antennas of the testbed's BS can be deployed flexibly, the KU Leuven Boardroom dataset provides three sets of individual sub-datasets with different antenna topologies, \textit{i}.\textit{e}, a uniform rectangular array (URA) of 8 by 8 antennas, a uniform linear array
(ULA) of 64 antennas, and a distributed antenna array (DIS). The details of the measurement environment are shown in~\cite{de2020csi}. 
%, and dichasus-0154 dataset\footnote{https://darus.uni-stuttgart.de/dataset.xhtml?persistentId=doi:10.18419/darus-2202}~\cite{dichasus2021,arnold2019novel}.
%\begin{itemize}
%\item{KU Leuven Boardroom dataset: }
%spatially labelled with an accuracy of less than 1 mm.   
%\item{Dichasus-0154 dataset: the dataset contains CSI and position tags of a massive MIMO-OFDM system measured by the DICHASUS channel sounder using a robot. The system has 32 antennas and 1024 sub-carriers, therefore the CSI at each postion has a shape of $(32,1024,2)$. The robot mostly stays in two rectangular areas inside the room. One area has a shape of $3$ m by $3$ m, and another $3$ m by $4$ m. The dataset is composed of $15233$ samples in total. We conduct two experiments on this dataset. One is based on the full dataset. For another, we resample the dataset to make sure the minimum distance between neighboring samples is larger than $0.1$ m, resulting in a smaller dataset containing 1262 samples. In both cases, the proportion of training samples, validation samples and test samples is set at 60\%, 20\% and 20\%, respectively.}
%\end{itemize}

 \subsubsection{Implementation Details}

\begin{itemize}
\item{Details of the network architecture: 
We first introduce the settings of the network's overall architecture. Specifically, we set $D=2$ and $p=q=2$ to reduce the spatial size of deep features by $1/16$. Then, we set $M=7$ and $z=1$ to further process the spatially-downsampled features. About the FCN, we set $n_{1}=64$ and $n_{2}=32$. The meaning of these parameters can be found in Section~\ref{archi}. As for the settings in the AAConv layer, we set $N_{q}=N_{v}=4$. Details about
these parameters are discussed in Section~\ref{sec aarb}.     
%For ULA and DIS situation in KU Leuven Boardroom dataset, where antennas are placed with a large distance, we set $N_{q}=N_{k}=N_{v}=N_{h}=4$, while for URA situation and Dichasus-0154 dataset, where antennas are closely placed, we set $N_{q}=N_{k}=N_{v}=1$, $N_{h}=4$.  
Besides, we use $5\times 5$ kernels for all Conv layers, the channel dimension of features is kept as $32$ for all blocks, and the activation layer we use is LeakyReLU.}
\item{Details on the training process: We set the batch size as $128$ and train the model for 1000 epochs with Adam optimizer. The initial learning rate is set as $10^{-3}$ for LOS datasets and $5\times10^{-4}$ for NLOS datasets, and decreases by half for every 200 epochs. We choose the model that performs the best on the validation set as the final model. All experiments run on a single GTX1080Ti GPU, and the codes are implemented by Tensorflow.}
\end{itemize}
\subsubsection{Comparison with State-of-the-art} We compare our methods with four SOTA CNN-based methods, whose training processes are set the same as ours for a fair comparison:
\begin{itemize}
\item{Sun19~\cite{sun2019fingerprint}: In~\cite{sun2019fingerprint}, a feature learning module, called CALP, is proposed, and the positioning method is composed of a cascade of five CALP modules followed by a FC layer.}
\item{Arnold19~\cite{arnold2019novel}: The method consists of two Conv layers, two pooling layers with a stride $(1,4)$ and four FC layers with 128 hidden neurons. For the Conv layers, we set the kernel size as $(5,5)$.}
\item{Chin20~\cite{chin2020intelligent}: The method has 
four cascaded gated Conv, Conv, and pooling layers with stride $(1,4)$, plus a $6$-layer FCN, where the hidden neurons have a configuration of $1024$-$512$-$256$-$256$-$64$-$2$. The method is originally proposed for a system with $924$ subcarriers; therefore, it uses a large pooling stride. When we apply it to the Ku Leuven Boardroom datasets, we reduce the pooling stride from $(1,4)$ to $(1,2)$. We also enlarge the convolution kernel from $(1,3)$ to $(5,5)$ for better performance.}
%two pooling layers with a stride $(1,4)$ and four FC layers with 128 hidden neurons.}
\item{Bast20~\cite{de2020csi}: The authors of~\cite{de2020csi} are also the releasers of Ku Leuven datasets. Their network contains 13 Conv layers, improved with skip connection and drop-out layers, and three FC layers at the end.}  
\item{{MHSA22~\cite{liu2022mhsa}: The authors of~\cite{liu2022mhsa}\footnote{The authors also propose a method called MHSA-EC by computing and using effective CSI. As the subcarrier used by each antenna is not provided in the dataset, we are unable to calculate the effective CSI, thus we do not compare with MHSA-EC.} also implement attention mechanism in their network. In their network, a 1D-CNN is used to extract features from the spatial dimension and the frequency dimension of CSI, respectively, followed by a stack of 1D self-attention layers. This setting differs from our method where these two dimensions are always jointly considered, where 2D-Conv and 2D-attention are utilized. We set the number of 1D Conv layers as $4$ and the feature maps across layers as $64$, which have been fine-tuned. Also, we use $4$ MHSA layers, following~\cite{liu2022mhsa}.}} 
\item{{Wang18~\cite{wang2018deep}: The positioning network used in~\cite{wang2018deep} contains $3$ Convs with $5\times 5$ kernels and $1$ Conv with $3\times 3$ kernels. A $(2,2)$ AveP is used after each Conv. At last, the features are flattened and mapped to locations via a FC layer. We use the raw CSI in KU Leuven datasets as the input to the network\footnote{The authors of~\cite{wang2018deep} also propose two preprocessing steps, 1) estimating AoA from CSI, and 2) constructing multiple images from AoA maps in different timestamps. These two steps cannot be repeated in Ku Leuven datasets, thus we omit these steps.}.}} 
\item{{Hoang20~\cite{hoang2020cnn}: Work~\cite{hoang2020cnn} uses three $5\times 5$ Convs with $10$ output features to extract information from CSI, followed by a 2-layer FCN. The number of hidden neurons in FCN is set as $1/10$ the feature size in CNN originally. This setting exceeds the GPU memory limit in the training phase. Thus, we reduce $1/10$ to $1/20$, and the FCN has a size of $3200$-$2$.}}
\end{itemize}
\subsubsection{Ablation Investigation}
To show the effectiveness of PBs and AARBs, we also present the performance of our network without PBs or AARBs. Specifically, we first replace the PBs and AARBs in our network with AvePs and RBs, respectively, and denote the resulting network as AAResCNN\_PB0\_AARB0. We then add PBs back and get AAResCNN\_AARB0. By comparing AAResCNN\_PB0\_AARB0 with AAResCNN\_AARB0 and comparing AAResCNN\_AARB0 and AAResCNN, we can prove the effectiveness of PBs and AARBs, respectively.
\subsubsection{Performance Evaluation}
\begin{table*}[t]
\caption{\label{table1} The MSE of different positioning methods in Ku Leuven boardroom datasets. The best results are shown in bold face}
\begin{center}
\resizebox{.92\columnwidth}{!}{ 
\begin{tabular}{|c|ccc|ccc|ccc|ccc|}
\hline
Dataset                 & \multicolumn{3}{c|}{\begin{tabular}[c]{@{}c@{}}Ku Leuven Boardroom\\ URA (mm)\end{tabular}} & \multicolumn{3}{c|}{\begin{tabular}[c]{@{}c@{}}Ku Leuven Boardroom\\ ULA (mm)\end{tabular}} & \multicolumn{3}{c|}{\begin{tabular}[c]{@{}c@{}}Ku Leuven Boardroom\\ DIS (mm)\end{tabular}} & \multicolumn{3}{c|}{\begin{tabular}[c]{@{}c@{}}Ku Leuven Lab\\ NLOS(mm)\end{tabular}}  \\ \hline
Training Sample $I$       & \multicolumn{1}{c|}{1000}          & \multicolumn{1}{c|}{5000}         & 10000         & \multicolumn{1}{c|}{1000}          & \multicolumn{1}{c|}{5000}         & 10000         & \multicolumn{1}{c|}{1000}          & \multicolumn{1}{c|}{5000}         & 10000        &\multicolumn{1}{c|}{1000}         &\multicolumn{1}{c|}{5000}  &10000  \\ \hline
Sun19                   & \multicolumn{1}{c|}{172.75}              & \multicolumn{1}{c|}{68.09}             &48.61               & \multicolumn{1}{c|}{423.73}              & \multicolumn{1}{c|}{107.33}             &47.81               & \multicolumn{1}{c|}{317.05}              & \multicolumn{1}{c|}{115.56}             &61.95                 &\multicolumn{1}{c|}{866.14}&\multicolumn{1}{c|}{533.94}&392.13  \\
Arnold19                & \multicolumn{1}{c|}{196.95}              & \multicolumn{1}{c|}{73.22}             &49.29               & \multicolumn{1}{c|}{205.97}              & \multicolumn{1}{c|}{62.94}             &40.84 & \multicolumn{1}{c|}{164.69}              & \multicolumn{1}{c|}{47.00}             &30.85               &\multicolumn{1}{c|}{785.42}&\multicolumn{1}{c|}{322.67}&204.54     \\
Chin20                  & \multicolumn{1}{c|}{148.53}              & \multicolumn{1}{c|}{64.49}             &31.80               & \multicolumn{1}{c|}{133.03}              & \multicolumn{1}{c|}{28.41}             &20.35               & \multicolumn{1}{c|}{99.00}              & \multicolumn{1}{c|}{26.23}             &15.21               &\multicolumn{1}{c|}{660.77}&\multicolumn{1}{c|}{248.25}&118.75      \\
Bast20                  & \multicolumn{1}{c|}{206.83}              & \multicolumn{1}{c|}{108.34}             &80.85 & \multicolumn{1}{c|}{275.32}              & \multicolumn{1}{c|}{89.22}             &42.98 & \multicolumn{1}{c|}{286.62}              & \multicolumn{1}{c|}{94.82}             &44.80
&\multicolumn{1}{c|}{826.42}                  
&\multicolumn{1}{c|}{483.28}
&356.61                    \\ 
MHSA22               & \multicolumn{1}{c|}{178.27}              & \multicolumn{1}{c|}{59.61}              &50.65 & \multicolumn{1}{c|}{291.74}              & \multicolumn{1}{c|}{81.90}             &41.55 &  \multicolumn{1}{c|}{282.40}              & \multicolumn{1}{c|}{91.13}              &50.65 &\multicolumn{1}{c|}{702.96}  &\multicolumn{1}{c|}{351.04}& 243.07  \\ 

Wang18               & \multicolumn{1}{c|}{336.42}              & \multicolumn{1}{c|}{239.48}              &185.60 & \multicolumn{1}{c|}{444.04}              & \multicolumn{1}{c|}{186.63}             &136.04 &  \multicolumn{1}{c|}{253.76}              & \multicolumn{1}{c|}{130.82}              &91.26 &\multicolumn{1}{c|}{1098.27}  &\multicolumn{1}{c|}{728.89}&626.95  \\ 

Hoang20               & \multicolumn{1}{c|}{242.30}              & \multicolumn{1}{c|}{98.44}              &68.96 & \multicolumn{1}{c|}{232.05}              & \multicolumn{1}{c|}{81.77}             &58.93 &  \multicolumn{1}{c|}{151.09}              & \multicolumn{1}{c|}{64.53}              &41.99 &\multicolumn{1}{c|}{820.75}  &\multicolumn{1}{c|}{463.94}& 342.99  \\ 

\hline
AAResCNN\_PB0\_AARB0(ours) & \multicolumn{1}{c|}{150.65}              & \multicolumn{1}{c|}{51.70}             &29.64               & \multicolumn{1}{c|}{95.96}              & \multicolumn{1}{c|}{26.86}             &18.04               & \multicolumn{1}{c|}{108.32}              & \multicolumn{1}{c|}{25.35}             &17.13          &\multicolumn{1}{c|}{725.03}&\multicolumn{1}{c|}{325.74}&151.00   \\
AAResCNN\_AARB0(ours)      & \multicolumn{1}{c|}{147.56}              & \multicolumn{1}{c|}{45.05}             &27.08 & \multicolumn{1}{c|}{79.53}              & \multicolumn{1}{c|}{26.36}             &17.61               & \multicolumn{1}{c|}{81.88}              & \multicolumn{1}{c|}{23.16}             &16.36   &\multicolumn{1}{c|}{666.06} &\multicolumn{1}{c|}{259.80}&131.94   \\
AAResCNN(ours)             & \multicolumn{1}{c|}{\textbf{124.63}}              & \multicolumn{1}{c|}{\textbf{42.86}}             &\textbf{26.15}  & \multicolumn{1}{c|}{\textbf{69.54}}              & \multicolumn{1}{c|}{\textbf{24.72}}             & \textbf{16.21}              & \multicolumn{1}{c|}{\textbf{67.26}}              & \multicolumn{1}{c|}{\textbf{21.97}}             & \textbf{14.17}                 &\multicolumn{1}{c|}{\textbf{598.96}}&\multicolumn{1}{c|}{\textbf{193.65}}&\textbf{108.34}  \\ \hline
\end{tabular}}
\end{center}
\end{table*}

Table~\ref{table1} compares the mean-squared error (MSE) between predicted locations $\ddot{l}$ and true locations $l$ for different positioning networks on the test datasets, which can be expressed as $\frac{1}{5000}\sum_{i=1}^{5000}\|\ddot{l^{i}}-l^{i}\|^2$. As we can see from Table~\ref{table1}, Chin20 performs the best among the existing methods, while the proposed AAResCNN improves the positioning accuracy of Chin20 by about 8\%-48\% in different cases, which denotes the superiority of our network in improving the positioning accuracy. In addition, we can see from the ablation investigation that both the PBs and AARBs are effective in decreasing the positioning error. 

\begin{figure}[t]
\centering
\subfloat[CDF versus MSE]{
\includegraphics[width=0.34\textwidth]{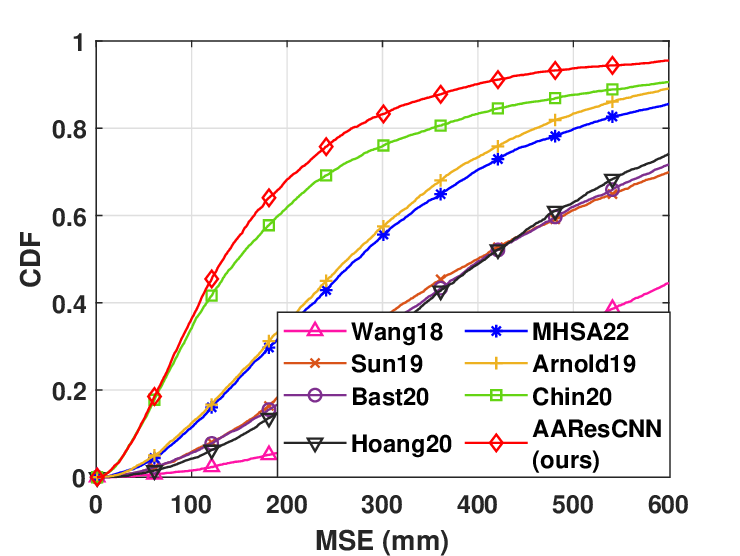}
}
\quad
\subfloat[Train/Val MSE versus Training Epoch]{
\includegraphics[width=0.32\textwidth]{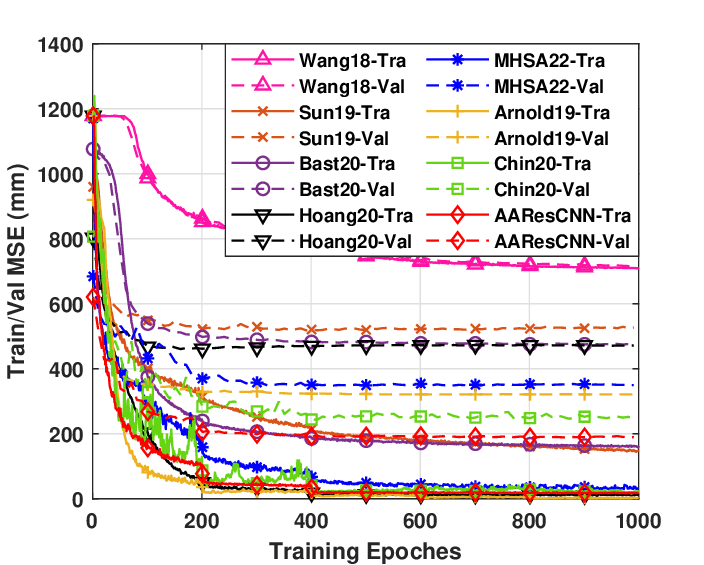}
}
\\
\caption{The other measurements of different positioning methods in the Lab NLOS dataset with $I=5000$ training samples. (a) CDF versus MSE (b) Train/Val MSE versus Training Epoch.}
\label{fig:position-cdf}
\end{figure}

Fig~\ref{fig:position-cdf}(a) shows the cumulative distribution function (CDF) of the positioning error using different methods in the Lab NLOS dataset. The results show that AAResCNN has a better localization performance than the SOTA methods. Fig~\ref{fig:position-cdf}(b) shows the training/validation losses changes over different training epochs. The results show that AAResCNN performs best on the validation datasets. 

\begin{figure}[t]
\centering
\subfloat[AARB1]{
\includegraphics[width=0.2\textwidth]{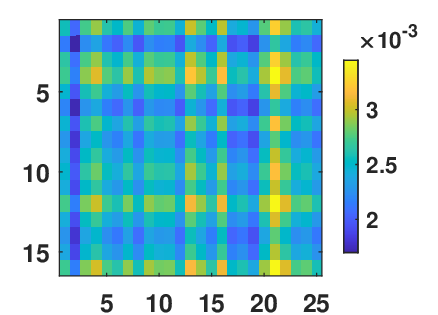}
}
\hfill
\subfloat[AARB2]{
\includegraphics[width=0.2\textwidth]{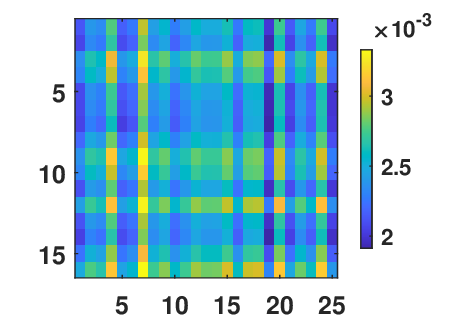}
}
\hfill
\subfloat[AARB3]{
\includegraphics[width=0.2\textwidth]{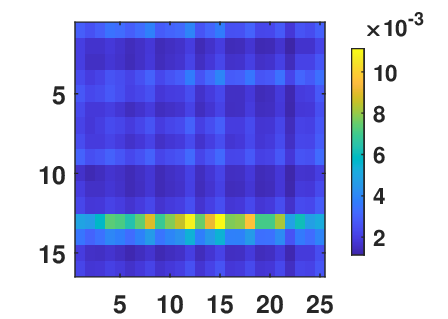}
}
\hfill
\subfloat[AARB4]{
\includegraphics[width=0.2\textwidth]{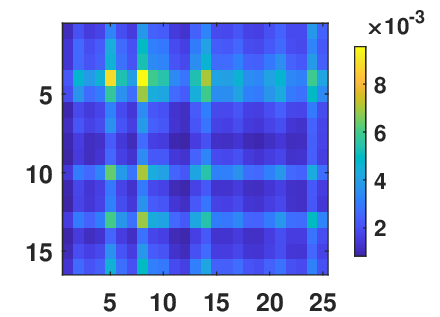}
}
\hfill
\subfloat[AARB5]{
\includegraphics[width=0.2\textwidth]{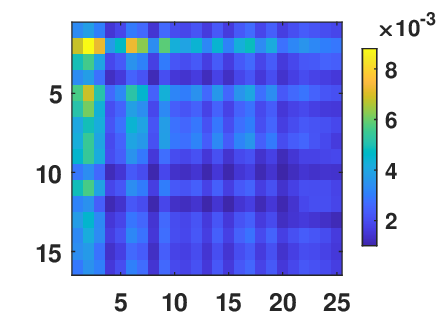}
}
\hfill
\subfloat[AARB6]{
\includegraphics[width=0.2\textwidth]{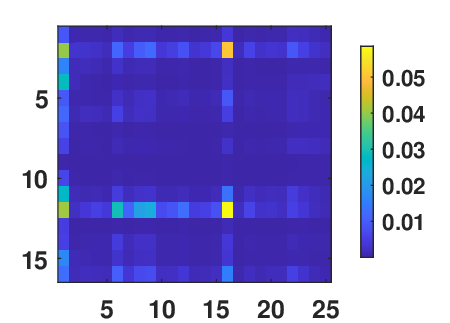}
}
\hfill
\subfloat[AARB7]{
\includegraphics[width=0.2\textwidth]{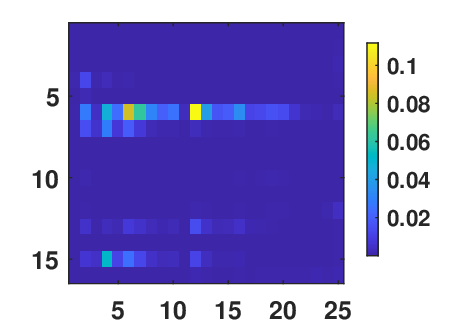}
}
\\
\caption{The visualization of the learned attention maps for pixel $(8,13)$ in the output features with spatial size $(16,25)$, \textit{i}.\textit{e}. $w^{O_{8,13}}$. (a)$\sim$(g) denote the attention maps from the 1st AARB to 7th AARB in the network. A brighter pixel in the attention map denotes a higher weight value.}
\label{fig:attention-map}
\end{figure}

Fig~\ref{fig:attention-map} visualizes the learned attention maps in different AARBs. We use one test sample in the ULA dataset and the network trained with $1000$ samples for visualization process. The learned attention maps are 4-D tensors with the shape $(16,25,16,25)$, and we focus on the maps for the central pixel $(8,13)$ in the output features for 2D demonstration, \textit{i}.\textit{e}. $w^{O_{8,13}}\in \mathcal{R}^{16\times 25}$. As shown in Fig~\ref{fig:attention-map}, the attention mechanism successfully captures longer range interactions, instead of focusing on a local region near $(8,13)$ or using each input pixel equally. Specially, the attention mechanism captures information from the whole input in the first two AARBs, with certain distances to avoid redundant information in neighboring carriers and antennas. As the network goes deeper, the attention mechanism gradually focuses more on several key points.
\begin{table}[t]
\caption{The model size and training/inference time of different methods}
\label{table3}
\centering
\resizebox{.7\columnwidth}{!}{ 
\begin{tabular}{|c|c|c|c|c|c|c|c|c|}
\hline
methods & Sun19 & Arnold19 & Chin20 & Bast20 &MHSA22 & Wang18 &Hoang20 &AAResCNN (ours)  \\ \hline
Size  & 7.7MB & 43.8MB   & 1.38GB  & 2.6MB &37.7MB &0.3MB&2.3GB  & 17.7MB \\
Training Time ($I=1000$)   &13.8min  &15.8min   &60.9min &21.3min &12.6min &12.7min &14.7min &      48.9min \\ 
Inference Time   &0.3ms  &0.2ms   &0.9ms    &0.2ms &0.3ms &0.2ms &0.2ms & 1.1ms       \\ \hline
\end{tabular}}
\end{table}

Table~\ref{table3} shows the model size and the training/inference time of different positioning networks. The model size is defined as the memory used to store the model and the inference time is the time used to predict the position from one CSI input. Compared with SOTA methods, our method has a slightly longer inference time, which, however, is acceptable considering the performance improvements. Besides, our method has an excellent parameter efficiency. Compared with Arnold19 and Chin20, we only use 40\% and 1.28\% memory to store our model, respectively.
\begin{table}[t]
\caption{\label{table4}The MSE, model size and speed of AAResCNN under varing values of parameters}
\centering
\resizebox{.9\columnwidth}{!}{ 
\begin{tabular}{|c|cccc|cccc|cccc|ccc|ccc|}
\hline
Param     & \multicolumn{4}{c|}{$M$}                                                                                   & \multicolumn{4}{c|}{$z$}                                                                                   & \multicolumn{4}{c|}{$N_{q}(=N_{v})$}                                                                       & \multicolumn{3}{c|}{$C_{in}(=C_{out})$}                                                                       
& \multicolumn{3}{c|}{$n_{1}(=2n_{2})$}                                                                        \\ \hline
value     & \multicolumn{1}{c|}{3}             & \multicolumn{1}{c|}{5}    & \multicolumn{1}{c|}{7}             & 8    & \multicolumn{1}{c|}{1}             & \multicolumn{1}{c|}{2}    & \multicolumn{1}{c|}{3}    & 4             & \multicolumn{1}{c|}{2}             & \multicolumn{1}{c|}{3}    & \multicolumn{1}{c|}{4}             & 5    & \multicolumn{1}{c|}{16}    & \multicolumn{1}{c|}{32}             & 64  & \multicolumn{1}{c|}{32}    & \multicolumn{1}{c|}{64}             & 128 \\ \hline
MSE (mm)   & \multicolumn{1}{c|}{101.8}         & \multicolumn{1}{c|}{86.3} & \multicolumn{1}{c|}{\textbf{69.5}} & 70.7 & \multicolumn{1}{c|}{\textbf{69.5}} & \multicolumn{1}{c|}{74.0} & \multicolumn{1}{c|}{77.8} & 80.1          & \multicolumn{1}{c|}{77.2}          & \multicolumn{1}{c|}{72.4} & \multicolumn{1}{c|}{\textbf{69.5}} & 69.7 &\multicolumn{1}{c|}{116.5} &\multicolumn{1}{c|}{\textbf{69.5}} & \textbf{69.5} &\multicolumn{1}{c|}{74.4} &\multicolumn{1}{c|}{\textbf{69.5}} &\multicolumn{1}{c|}{71.9}\\ \hline
Size (MB)  & \multicolumn{1}{c|}{\textbf{15.0}} & \multicolumn{1}{c|}{16.4} & \multicolumn{1}{c|}{17.7}          & 18.4 & \multicolumn{1}{c|}{17.7}          & \multicolumn{1}{c|}{17.5} & \multicolumn{1}{c|}{17.4} & \textbf{17.4} & \multicolumn{1}{c|}{\textbf{17.6}} & \multicolumn{1}{c|}{17.6} & \multicolumn{1}{c|}{17.7}          & 17.8 &\multicolumn{1}{c|}{\textbf{7.2}} &\multicolumn{1}{c|}{17.7} &\multicolumn{1}{c|}{50.3} &\multicolumn{1}{c|}{\textbf{13.0}} &\multicolumn{1}{c|}{17.7} &27.2\\ \hline
Time (ms) & \multicolumn{1}{c|}{\textbf{0.6}}  & \multicolumn{1}{c|}{0.9}  & \multicolumn{1}{c|}{1.1}           & 1.2  & \multicolumn{1}{c|}{1.1}           & \multicolumn{1}{c|}{0.8}  & \multicolumn{1}{c|}{0.7}  & \textbf{0.7}  & \multicolumn{1}{c|}{\textbf{1.0}}  & \multicolumn{1}{c|}{1.0}  & \multicolumn{1}{c|}{1.1}           & 1.1  &\multicolumn{1}{c|}{\textbf{0.9}} &\multicolumn{1}{c|}{1.1} &1.3 &\multicolumn{1}{c|}{\textbf{1.1}} &\multicolumn{1}{c|}{1.1} &1.1\\ \hline
\end{tabular}}
\end{table} 
\subsubsection{Parameter Analysis}
In the above experiments, we set $M=7$, $z=1$ and $N_{q}=N_{v}=4$ by default. Here we explain our settings through comparative experiments, where the value of one parameter changes while the others keep constant. The experiments are conducted on the ULA dataset with $I=1,000$ training samples. The results are shown in Table~\ref{table4}. As we can see from Table~\ref{table4}, the defaulted settings have the best trade-off between performance and cost. For example, when $M\leq7$, the MSE falls as the value of $M$ increases, while the MSE stops decreasing when $M\geq 7$. At the same time, the running time and the model size keeps increasing as $M$ grows. Therefore, we set $M=7$.

\subsubsection{Positioning error distribution}
\label{Error-distribution}
\begin{figure}[t]
\centering
\subfloat[URA]{
\includegraphics[width=0.25\textwidth]{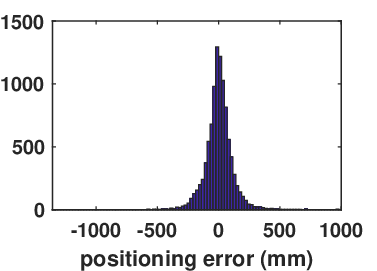}
}
\hfill
\subfloat[ULA]{
\includegraphics[width=0.25\textwidth]{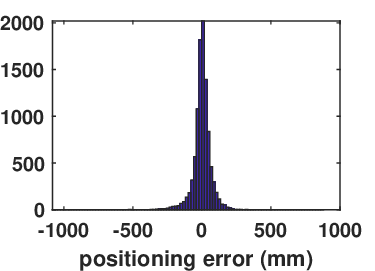}
}
\hfill
\subfloat[DIS]{
\includegraphics[width=0.25\textwidth]{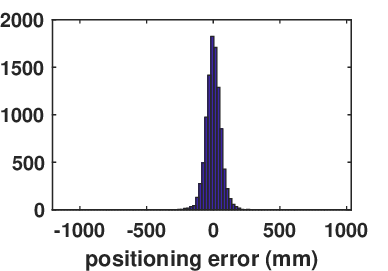}
}
\\
\caption{The positioning error distributions of the learned AAResCNNs on three different test datasets, \textit{i}.\textit{e}. the URA, ULA, DIS sub-dataset in KU Leuven boardroom dataset. The AAResCNN is trained with $1,000$ samples in each dataset. The positioning error calculated for the $x$ dimension and $y$ dimension is shown together in the histograms.}
\label{fig:position-error-dis}
\end{figure}
Finally, we show the positioning error distributions created by the learned AAResCNNs in Fig~\ref{fig:position-error-dis}. As shown in Fig~\ref{fig:position-error-dis}, the measurement noises of different AAResCNNs are Gaussian noises with zero mean and various standard deviations, which verifies the assumption used in Eq (\ref{equ:t4}). 
\subsection{Tracking across environments}
\label{exp: track}
In this subsection, we will compare the generality of different tracking systems across different CSI environments. Specifically, we first generate $N_{L}$ training/validation trajectories in the measurement areas of Ku Leuven datasets and collect the CSI along the trajectories in boardroom URA dataset. This CSI-trajectory dataset is then used to design different tracking systems. After this, we generate test CSI-trajectory datasets in boardroom URA/ULA/DIS and lab LOS datasets\footnote{\url{https://homes.esat.kuleuven.be/~sdebast/measurements/measurements_lab.html}}, and evaluate the performance of the pre-trained tracking systems directly without retraining. In other datasets except for the URA dataset, a positioning system trained with $I$ reference points is available. As collecting a CSI-trajectory dataset is more time-consuming than a CSI-location dataset, it is imperative for the tracking system designed under one environment to work for environments where trajectory datasets are unavailable.  
\subsubsection{Trajectory Datasets}
We consider both the synthesized and the real trajectory dataset. For the synthesized dataset, we consider the following three motion patterns: (a) walking with a constant velocity; (b) walking straight with arbitrary velocity; (c) changing direction at a certain step based on pattern b. The three types arise with equal probability and each trajectory sample consists of $T=5$ steps. We generate $8,000$ for training, $2,000$ for validation, and $1,000$ for test. For the real trajectory dataset, we use publicly-released ReSysTDepth indoor trajectory dataset\footnote{\url{https://ieee-dataport.org/documents/resystdepth-real-and-synthetic-trajectories-indoor-environments-captured-depth-sensors}}. We rescale the trajectories inside to a $1.25$ m by $1.25$ m area and find their closest mappings in the left-upper grid of the measurement areas in KU Leuven dataset. We divide each trajectory into $T=10$ sub-trajectories, resulting in $4000$ trajectories. We use $2,400$ for training, $600$ for validation, and $1,000$ for test.
\subsubsection{Implementation Details of our Methods and State-of-the-art}
%According to whether the CSI-based positioning network is required during the training phase, we divide the compared methods into two categories: CSI-dependent and CSI-free. %For CSI-dependent methods, we train their tracking methods using the positioning network trained in Ku Leuven boardroom URA dataset with 5000 samples. For CSI-free methods, the networks are trained from the trajectory
In this section, we describe the training and testing details of the compared methods. 
\begin{itemize}
\item{AAResCNN: we train an AAResCNN for each environment using all the available training samples. Specifically, for the boardroom URA dataset, we divide the $N_{L}$ $T$-step trajectories into $N_{L}T$ samples and train an AAResCNN, while for the others, we use the $I$ samples for training. During testing, each point along the trajectory is recovered independently using the trained AAResCNNs.}
%\item{Kalman filter: following~\cite{bai2019dl}, we compare with kalman-filter-based methods. We set the state as $[l_{t},l_{t-1},l_{t-2}]$, and predicts $l_{t+1}$ from current state by $l_{t+1}=l_{t}+1/2*(l_{t}-l_{t-2})$. And $\tilde{l}_{t}$ is adopted as measurements at time $t$.}
\item{LSTM-F: we compare with the LSTM-based tracking method proposed in~\cite{hoang2020cnn,zhang2021deep}. Following~\cite{hoang2020cnn}, we first train an AAResCNN using all the training samples. And then, we input the CSI to the pre-trained AAResCNN and extract the deep features before the last FC layer. Finally, an LSTM network is used to transfer the deep features of $T$ time steps into locations. During training, we use the AAResCNN trained under boardroom URA dataset to extract deep features and train the LSTM network. During testing, we reuse the LSTM network and extract deep features using the AAResCNN trained for each environment.}
\item{LSTM-P: the method is similar to LSTM-F; the only difference is that we use the predicted locations of the AAResCNN as the inputs of the LSTM network.}
%\item{LSTM-P: the method is similar to LSTM-F, the only difference is that we use the predicted locations of the pre-trained positioning network as the inputs for the LSTM.}
\item{DNN-prior (ours): We first train an AAResCNN for each environment. Then we implement our prior-based tracking method using $N_{L}$ trajectories in the boardroom URA dataset, following Section~\ref{sec:track process}. The Gaussian Denoisers are trained under $\sqrt{\lambda}=1$, $3$, $[5,70]$ with an interval of $5$, respectively. Finally, the tracking system is used on top of the AAResCNN in each environment by tuning the value of $\lambda$.}
\end{itemize}
Note that we do not compare with Bayesian-filtering-based method here since the MT has arbitrary velocity in our experiments, making it hard to define system models.
\subsubsection{Performance Evaluation}
\begin{table}[t]
\begin{center}
\caption{\label{table2} The MSE of different tracking methods across environment. The best results are shown in bold face.}
\resizebox{1.\columnwidth}{!}{ 
\begin{tabular}{|c|c|ccc|ccc|ccc|}
\hline
Synthesized Trajectory Dataset      & Boardroom URA (mm)  & \multicolumn{3}{c|}{Boardroom ULA (mm)}                                 & \multicolumn{3}{c|}{Boardroom DIS (mm)}                                  & \multicolumn{3}{c|}{Lab LOS (mm)}                                       \\ \hline
Training Sample & $N_{L}$=10000       & \multicolumn{1}{c|}{I=500} & \multicolumn{1}{c|}{I=1000} & I=5000 & \multicolumn{1}{c|}{I=500} & \multicolumn{1}{c|}{I=1000} & I=5000 & \multicolumn{1}{c|}{I=1000} & \multicolumn{1}{l|}{I=5000} & I=10000 \\ \hline
AAResCNN        &8.90               & \multicolumn{1}{c|}{164.70}       & \multicolumn{1}{c|}{68.32}       &24.66         & \multicolumn{1}{c|}{155.41}       & \multicolumn{1}{c|}{67.05}       &21.58        & \multicolumn{1}{c|}{172.50}       & 
\multicolumn{1}{c|}{75.46}       &46.70        \\
%Kalman-filter        &\textbf{}               & \multicolumn{1}{c|}{}       & \multicolumn{1}{c|}{}       & & \multicolumn{1}{c|}{}       & \multicolumn{1}{c|}{}       &        & \multicolumn{1}{c|}{}       & \multicolumn{1}{c|}{}       &         \\
LSTM-F        &\textbf{8.30}               & \multicolumn{1}{c|}{1044.36}       & \multicolumn{1}{c|}{1111.20}       &1065.30 & \multicolumn{1}{c|}{998.31}       & \multicolumn{1}{c|}{1356.26}       &1232.88        & \multicolumn{1}{c|}{1099.45}       & \multicolumn{1}{c|}{1315.23}       &1431.48         \\
LSTM-P       &8.42               & \multicolumn{1}{c|}{164.76}       & \multicolumn{1}{c|}{68.36}       &24.74        & \multicolumn{1}{c|}{155.62}       & \multicolumn{1}{c|}{67.11}       &21.72 & \multicolumn{1}{c|}{172.50}       & \multicolumn{1}{c|}{75.51}       &46.80        \\
DNN-prior (ours) &8.99               & \multicolumn{1}{c|}{\textbf{145.00}}       & \multicolumn{1}{c|}{\textbf{63.71}}       &\textbf{24.17}         & \multicolumn{1}{c|}{\textbf{130.26}}       & \multicolumn{1}{c|}{\textbf{60.64}}       &\textbf{21.33}         & \multicolumn{1}{c|}{\textbf{146.86}}       & \multicolumn{1}{c|}{\textbf{68.76}}       &\textbf{43.06}         \\ \hline

Real Trajectory Dataset        & Boardroom URA (mm)  & \multicolumn{3}{c|}{Boardroom ULA (mm)}                                 & \multicolumn{3}{c|}{Boardroom DIS (mm)}                                  & \multicolumn{3}{c|}{Lab LOS (mm)}                                                                           \\ \hline
Training Sample &$N_{L}$=3000       & \multicolumn{1}{c|}{I=500} & \multicolumn{1}{c|}{I=1000} &I=5000 & \multicolumn{1}{c|}{I=500} & \multicolumn{1}{c|}{I=1000} & I=5000 & \multicolumn{1}{c|}{I=1000} & \multicolumn{1}{l|}{I=5000} &I=10000 \\ \hline
AAResCNN        &3.44              & \multicolumn{1}{c|}{98.82}       & \multicolumn{1}{c|}{40.13}       &16.71        & \multicolumn{1}{c|}{140.55}       & \multicolumn{1}{c|}{56.99}       &20.17        & \multicolumn{1}{c|}{109.82}       & \multicolumn{1}{c|}{60.20}       &34.50       \\
%Kalman-filter        &\textbf{}               & \multicolumn{1}{c|}{}       & \multicolumn{1}{c|}{}       & & \multicolumn{1}{c|}{}       & \multicolumn{1}{c|}{}       &        & \multicolumn{1}{c|}{}       & \multicolumn{1}{c|}{}       &         \\
LSTM-F        &\textbf{3.33}               & \multicolumn{1}{c|}{1085.12}       & \multicolumn{1}{c|}{1118.85}       &398.61 & \multicolumn{1}{c|}{672.91}       & \multicolumn{1}{c|}{1020.35}       &1208.48        & \multicolumn{1}{c|}{1253.99}       & \multicolumn{1}{c|}{1385.74}       &1267.87         \\
LSTM-P       &3.37             & \multicolumn{1}{c|}{98.58}       & \multicolumn{1}{c|}{39.90}       &16.31       & \multicolumn{1}{c|}{139.82}       & \multicolumn{1}{c|}{56.50}       &19.99 & \multicolumn{1}{c|}{109.73}       & \multicolumn{1}{c|}{59.98}       &34.27        \\
DNN-prior (ours) &4.08              & \multicolumn{1}{c|}{\textbf{67.81}}       & \multicolumn{1}{c|}{\textbf{32.21}}       &\textbf{14.96}         & \multicolumn{1}{c|}{\textbf{99.83}}       & \multicolumn{1}{c|}{\textbf{44.28}}       &\textbf{17.94}         & \multicolumn{1}{c|}{\textbf{77.84}}       & \multicolumn{1}{c|}{\textbf{43.11}}       &\textbf{27.48}         \\ \hline
\end{tabular}}
\end{center}
\end{table}

We compare the performance in Table~\ref{table2}. The performance is evaluated by the MSE between the last step of refined trajectory $\hat{L}$ and true trajectory $L$, denoted as $\frac{1}{1000}\sum_{i=1}^{1000}\|\hat{L}^{i}_{T}-L^{i}_{T}\|^2$. As shown in Table~\ref{table2}, despite the fact that LSTM-F performs the best in the URA dataset, the method can not be reused for any other environment, as the feature extraction process in each AAResCNN differs significantly. The performance of LSTM-P is not better than AAResCNN in any considered dataset other than the URA dataset, although it can be applied across environments, which means it fails to utilize historical trajectory information effectively when the test environment is different from the training one. On the contrary, the proposed prior-based method works for any environment. It improves the performance of AAResCNN by about 1\%-31\% depending on the accuracy of AAResCNN except for the boardroom URA dataset, where AAResCNN itself is already highly-accurate and our method becomes ineffective.
\subsection{IMU-aided Tracking With Varying IMU Precision}
Here, we will focus on IMU-aided tracking systems and verify the superiority of our plug-and-play based method in dealing with IMU measurements with different precisions. We only consider the Boardroom ULA dataset and use the same test dataset generated in Section~\ref{exp: track}. As for the IMU measurements, we first calculate the true step length $r_{t}$ and orientation $\theta_{t}$ from the trajectories. Then we add noise to them to satisfy different SNR requirements according to~(\ref{equ:t18}) and~(\ref{equ:t19}). In addition, we collect $I$ CSI-location samples to train the positioning network. We also generate $10,000$ trajectories for training purposes and the CSI along the training trajectories is considered to be unknown.
\subsubsection{Implementation Details of our Methods and State-of-the-art}
Besides comparing with the positioning network AAResCNN and the IMU-free tracking method DNN-prior, we also consider the following IMU-aided tracking methods: 
\begin{itemize}
\item{Particle Filter~\cite{belmonte2019swiblux}: In~\cite{belmonte2019swiblux}, a particle filter based method has been proposed to fuse the fingerprinting-based positioning results and IMU measurements. We reimplement this method for comparison.}
\item{DNN-SNR1-100: this method is a data-driven extension of our prior-based tracking approach to IMU-aided tracking. Besides using the noisy locations (training phase) or predicted results from a positioning network (testing phase) as the inputs of the DNN, we also feed the IMU measurements and the corresponding SNR value into the network. The method is trained with IMU measurements from SNR=1 dB to SNR=100 dB simultaneously. We also fine-tune the noise level added during the training phase.}
\item{DNN-SNR1 and DNN-SNR100: Different from DNN-SNR1-100, the methods only consider SNR=1 dB and 100 dB during training, respectively.}
\item{PnP (ours): the details of our PnP-based IMU-aided tracking method are described in Section~\ref{imu tracking}. We use the Gaussian Denoiser that performs the best in Section~\ref{exp: track} under the same testing environment. For IMU measurements with different SNR values, we fine-tune the values of $\mu$ and $\rho$.} %We tune the trade-off parameters for varying SNR.}
\end{itemize}
\subsubsection{Performance Evaluation}
\begin{figure}[t]
\centering
\subfloat[ULA I=1000]{
\includegraphics[width=0.35\textwidth]{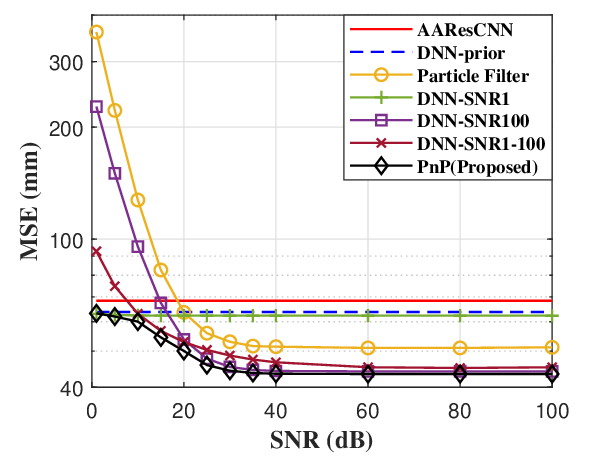}
}
\quad
\subfloat[ULA I=5000]{
\includegraphics[width=0.35\textwidth]{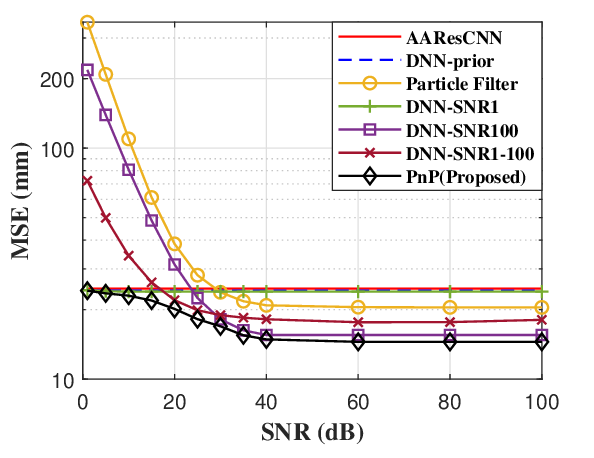}
}
\\
\caption{The performance comparison of different IMU-aided tracking methods in the boardroom ULA dataset when the precision of IMU measurements changes. As the number of training samples affects the performance, we consider two situations: (a) $I=1000$ CSI-position training samples are collected; (b) $I=5000$ CSI-position training samples are collected.}
\label{fig:track}
\end{figure}

\begin{figure}[t]
\centering
\subfloat[Trajectory 1]{
\includegraphics[width=0.3\textwidth]{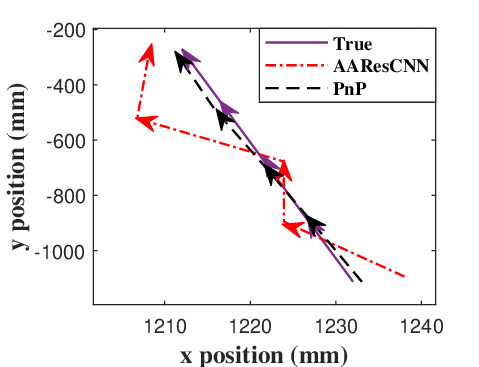}
}
\hfill
\subfloat[Trajectory 2]{
\includegraphics[width=0.3\textwidth]{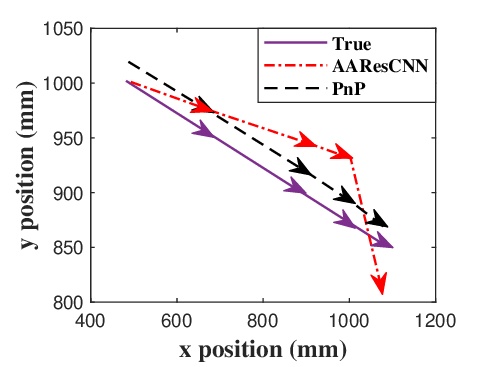}
}
\hfill
\subfloat[Trajectory 3]{
\includegraphics[width=0.3\textwidth]{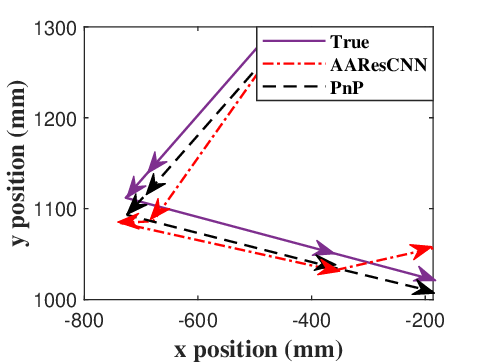}
}
\\
\caption{Some examples of true trajectories (purple lines), recovered trajectories via positioning network AAResCNN (red lines) and refined trajectories by IMU-aided tracking method PnP (black lines).}
\label{fig:examples}
\end{figure}
We compare the performance of different IMU-aided tracking methods in Fig~\ref{fig:track}, where the MSE vs. SNR is reported. As shown in Fig~\ref{fig:track}, when the SNR value of IMU measurements is high, all the IMU-aided tracking methods achieve better performance than AAResCNN and DNN-prior. However, these methods perform differently when the SNR value changes. For the methods that assume IMU measurements are highly-reliable during the deployment stage, \textit{e}.\textit{g}., particle filter and DNN-SNR100, the performance degrades significantly when the precision of IMU equipment is low in the test phase. While for the method DNN-SNR1, which is trained based on the low SNR value, its performance gets no improvement when a precise IMU equipment is used. Also, for method DNN-SRN1-100, which takes SNR value as input and considers SNR changes during the training stages, it raises the localization error of DNN-SNR100 in the high SNR range, although it performs better than DNN-SNR100 in the low SNR range. Compared to these methods, our PnP-based IMU-aided tracking method has the best performance in most cases, which signifies its flexibility in dealing with precision difference of IMU. 

To show the functionality of IMU measurements in reducing positioning error, we also show some examples of true trajectories, recovered trajectories by positioning network AAResCNN and refined trajectories by IMU-aided tracking method PnP in Fig~\ref{fig:examples}. The examples are extracted under the ULA dataset with $I=5,000$ training samples. The SNR of IMU measurements is set as 100. As shown in Fig~\ref{fig:examples}, the refined trajectories are closer to the true trajectories. 
\section{Conclusion}
\label{conclusion}
In this paper, we have investigated CSI-based fingerprinting indoor positioning and tracking for MIMO-OFDM systems. First, we have proposed a new positioning network called AAResCNN, which achieves the lowest positioning error compared with SOTA CNN-based positioning approaches. And then, we improve the generality of tracking and IMU-aided tracking methods. Based on the deep trajectory prior and the idea of plug-and-play, the proposed tracking method can be applied to any wireless environment and is compatible with IMU equipments with arbitrary precision. Numerical results on publicly-released datasets demonstrate the effectiveness of our methods.
%\newpage
%\small
\bibliographystyle{IEEEtran}
\bibliography{ref}
\small
\end{document}